\documentclass[fleqn,usenatbib]{mnras}

\usepackage{newtxtext,newtxmath}

\usepackage[T1]{fontenc}
\usepackage{ae,aecompl}

\usepackage{graphicx}	
\usepackage{amsmath}	
\usepackage{amssymb}	

\title[Bow shock polarization -- electrons]{\textit{Polarization simulations of stellar wind bow shock nebulae. I. The case of electron scattering}}

\author[M. Shrestha et al.]{
Manisha Shrestha,$^{1}$\thanks{E-mail: manisha.shrestha9@du.edu (DU)}
Hilding R. Neilson,$^{2}$
Jennifer L. Hoffman,$^{1}$
Richard Ignace$^{3}$
\\
$^{1}$Department of Physics and Astronomy, University of Denver, 2112 E. Wesley Ave., Denver, CO, 80208, USA\\
$^{2}$Department of Astronomy and Astrophysics, University of Toronto, 50 St. George Street, Toronto, M5S 3H4, Canada\\
$^{3}$Department of Physics and Astronomy, East Tennessee State University, Johnson City, TN, 37614, USA
}
\date{Last updated 2016 June 27; in original 2015}

\pubyear{2017}

\date{Accepted XXX. Received YYY; in original form ZZZ}

\pubyear{2015}

\begin{document}
\label{firstpage}
\pagerange{\pageref{firstpage}--\pageref{lastpage}}
\maketitle

\begin{abstract}

Bow shocks and related density enhancements produced by the winds of massive stars moving through the interstellar medium provide important information regarding the motions of the stars, the properties of their stellar winds, and the characteristics of the local medium. Since bow shock nebulae are aspherical structures, light scattering within them produces a net polarization signal even if the region is spatially unresolved. Scattering opacity arising from free electrons and dust leads to a distribution of polarized intensity across the bow shock structure. That polarization encodes information about the shape, composition, opacity, density, and ionizsation state of the material within the structure. In this paper we use the Monte Carlo radiative transfer code \textit{SLIP} to investigate the polarization created when photons scatter in a bow shock-shaped region of enhanced density surrounding a stellar source. We present results for electron scattering, and investigate the polarization behaviour as a function of optical depth, temperature, and source of photons for two different cases: pure scattering and scattering with absorption. In both regimes we consider resolved and unresolved cases. We discuss the implications of these results as well as their possible use along with observational data to constrain the properties of observed bow shock systems. In different situations and under certain assumptions, our simulations can constrain viewing angle, optical depth and temperature of the scattering region, and the relative luminosities of the star and shock.

\end{abstract}

\newpage



\section{Introduction} \label{intro}
Mass loss from massive stars impacts their evolution  \citep[e.g.,][]{2012ARA&A..50..107L} as well as the evolution and dynamics of the surrounding interstellar medium  \citep[ISM;][]{1975ApJ...200L.107C}. One of the most visible manifestations of stellar mass loss, a bow shock, forms when the stellar wind emanating from a star moving through the ISM reaches supersonic relative velocities \citep[e.g.,][]{Wilkin_1996}. The properties of such stellar wind bow shocks encode information about the mass-loss history of the star  \citep[e.g.,][]{2008ApJ...685L.141R,Mackey_Mohamed_Neilson_Langer_Meyer_2012,Gvaramadze_Menten_Kniazev_Langer_Mackey_Kraus_Meyer_Kami?ski_2013} and the structure of the surrounding ISM \citep[e.g.,][]{2011ApJ...737..100T}. 

Most observed bow shocks are associated with massive runaway stars; however, they are also observed around a variety of stellar sources including asymptotic giant branch stars \citep[e.g.,][]{Ueta_2006}, pulsars \citep[e.g.,][]{cordes_romani_1996}, cataclysmic variables \citep[e.g.,][]{Buren_93}, and Algols \citep[e.g.,][]{Mayer_2016}. These bow shocks are typically detected at optical \citep[e.g.,][]{gull_1979} and infrared (IR) wavelengths  \citep[e.g.,][]{buren_mccray_1988,Ueta_2006,Ueta_Izumiura_Yamamura_Nakada_Matsuura_Ita_Tanab_Fukushi_2014}, though a few have been detected at X-ray \citep[e.g.,][]{Lopez_2012}, ultraviolet \citep[e.g.,][]{LeBertre_2012}, and radio \citep[e.g.,][]{Benaglia_2010} wavelengths.  In recent years, several dedicated surveys have revealed large numbers of bow shock nebulae in the Milky Way \citep[e.g.,][]{Peri_Benaglia_Brookes_Stevens_Isequilla_2011,Peri_2015,Kobulnicky_2016}, opening new avenues of research into stellar winds and ISM characteristics.

In this paper, we probe the connections between polarimetric observations and the physics of stellar wind bow shocks. (Hereafter, we will use the term ``bow shock" to describe not only a true physical shock, but also a region of enhanced density arising from wind-ISM interactions and having the same geometrical shape as a bow shock.) Polarization by scattering samples the opacity of a medium, and encodes information about the relative orientation of a scattering region in relation to illuminating sources and the observer. In the case of electron (Thomson) scattering, interaction of unpolarized incident radiation with a free electron produces scattered radiation that is $100\%$ linearly polarized when the scattering angle is $90\degr$, independent of wavelength; the angle of polarization is perpendicular to the plane defined by the incident and scattered rays \citep{rybicki_1979}. In the case of dust scattering, asymmetric dust grains produce scattered radiation whose linear polarization magnitude and position angle are wavelength-dependent, and which may also be circularly polarized \citep{Henyey_1941,White_1979}. Polarization has been detected in two bow shock sources near the Galactic centre, with magnitudes up to a few percent \citep{Buchholz_2011,2013A&A...551A..35R}. Such values are easily measured with current polarimetric instrumentation, suggesting that polarization may be a valuable technique with which to study the wealth of newly discovered bow shocks. 

Although many researchers have developed computational models of stellar wind bow shocks \citep[e.g., ][]{Gustafsson_2010,Mohamed_2013,Christie_2016}, polarization signatures have not generally been considered. However, a few recent papers have modelled the polarization of specific objects with bow shocks. \cite{2013msao.confE.172N}  analytically modelled the near-IR polarization from a bow shock around Betelgeuse.  \citealt{Shahzamanian_2016} and \citealt{zajacek_2017} used a sophisticated 3-D Monte Carlo radiative transfer (MCRT) code to simulate the polarization behaviour of a dust-scattering bow shock and other possible circumstellar structures around the Dusty S-cluster Object (DSO), an unusual infrared-excess source near the Galactic centre.

This contribution is the first of two papers in which we use Monte Carlo numerical methods to explore the polarization signatures arising from generalized stellar wind bow shock structures. Our code (\textit{SLIP}; \citealt{Hoffman_2007}) is related to the one used by  \citealt{Shahzamanian_2016} and \citealt{zajacek_2017}, but our implementation is different, as discussed below in Section~\ref{methods}. The MCRT approach is easily adaptable to non-spherical geometries while allowing for consideration of optical depth effects (i.e., the influence of multiple scattering on the polarization of escaping light). Our goal in this  paper is to formulate the problem of predicting the polarization produced within an idealized bow shock structure and to investigate the effects of various input parameters on the resulting polarization behavior, assuming Thomson scattering only for simplicity.  The second paper (hereafter Paper~II) will investigate the effects of dust opacity on observed polarization, a scenario with broader applications.

Our paper is organized as follows. In Section \ref{methods}, we discuss the \textit{SLIP} code and the features of our models. In Section \ref{analytic}, we present analytic results for our idealized bow shock cases, valid strictly in the optically thin limit. Although limited in applicability, the analytic results provide context for interpreting the numerical results from \textit{SLIP}. In this section we also discuss comparisons between the analytic and numerical simulations. In Section \ref{results}, we present and interpret numerical results for the polarization produced in both resolved and unresolved cases, as functions of the temperature and optical depth of the scattering material in the bow shock. We discuss how our results may aid in interpretation of observed polarization signals in Section \ref{obsimp}. Finally, we offer concluding remarks in Section \ref{conclusion}.
 
\section{Methods}
\label{methods}

We constructed our simulations using the Supernova LIne Profile (\textit{SLIP}) code (\citealt{Hoffman_2007}). \textit{SLIP} uses the MCRT method (e.g., \citealt{Whitney_2011}) to track photons through a three-dimensional spherical polar grid as in \cite{Whitney_Wolff_2002}. For the axisymmetric simulations presented here, we define a grid with $100$ radial cells and $101$ cells in the polar ($\theta$) direction.

At the centre of this grid we place a finite spherical photon source, surrounded by a circumstellar scattering region composed of pure hydrogen in local thermodynamic equilibrium (LTE). We do not assume this circumstellar material (CSM) is heated by the central star. Instead we define its temperature $T$ (which for simplicity we assume is constant throughout the region) as a user-specified input parameter governing the ionization fraction $x$ within the scattering region. Given a specified reference optical depth $\tau_0$, \textit{SLIP} first calculates the number density of free electrons via the equation~$n_+ = \tau_0/0.4m_H\Delta R_0$, where $m_H$ is the proton mass and $\Delta R_0$ is the radial thickness of the scattering region at the reference location. These quantities are defined in greater detail later in this section. With this value of $n_+$ and the input temperature $T$, we then apply the Saha equation to calculate $n_0$, the number density of neutral atoms: 

\begin{equation}
\frac{n_+}{n_0} = \frac{Z_+}{Z_0} \frac{2}{n_e h^3} (2 \pi m_e k T)^{\frac{3}{2}} e^{\frac{-\chi_i}{kT}}
\label{saha}
\end{equation}

\noindent In this equation, $n_e$ represents the number density of free electrons, $m_e$ the electron mass, and $k$ the Boltzman constant. $Z_+$ and $Z_0$ represent the partition functions of the ion and neutral atom, respectively, and $\chi$ is the ionization potential. From the calculated $n_0$ value, we obtain the ionization fraction $x=n_+/n_{tot}$ and finally the opacity of the CSM, $\kappa=0.4x$. By doing this, we assume a constant ionization fraction and opacity throughout the CSM, which simplifies the Monte Carlo calculations described below. The code does not take into account any expansion of the CSM, which is a reasonable approximation for the case of a roughly stationary stellar wind bow shock. 

Following the basic MCRT prescription, \textit{SLIP} emits virtual, initially unpolarized ``photons" from the central star (or other photon source) and tracks them as they scatter within the CSM. The code determines a photon's behaviour by generating weighted random numbers corresponding to known probability distributions that depend on the optical depth $\tau$ and albedo $a$ of the scattering region  \citep{Whitney_2011}. A strength of our implementation is that in addition to the star (or ``central source"), \textit{SLIP} also allows photons to be emitted from within the CSM itself (which we refer to as the ``distributed source"). In the distributed emission case, we allow photons to be emitted isotropically from the volume of the CSM. Because the CSM density is not constant (see the discussion of the bow shock implementation below), we use the rejection method to ensure that the number of emitted photons at a given location is proportional to the local CSM density. In the sections below, we investigate the differences between these two emission scenarios. 

As photons interact with the scattering region, \textit{SLIP} performs the numerical optical depth integration described in \citet{Code_Whitney_1995} and \citet{Whitney_2011}. After each integration, a random number compared with the photon's albedo determines whether it scatters or becomes absorbed; the photon's Stokes parameters are updated after each scattering event by applying the standard Mueller matrix multiplication \citep{Chandrasekhar_1946,Code_Whitney_1995,Whitney_2011}. Once a photon exits the simulation (i.e., it ``escapes''), its Stokes parameters are combined with those of all previously tracked photons in the appropriate output bin corresponding to the observer's viewing angle. A single \textit{SLIP} run produces results for all viewing angles ($i=0\degr-180\degr$). Within each output bin, we sum the Stokes vectors due to all $N$ photons in the bin and apply normalization factors in $\theta$ and $\phi$ to ensure that output fluxes have the correct units. We determine the $1\sigma$ uncertainty for each Stokes parameter in each bin by calculating the standard deviation of that parameter over all $N$ photons in the bin and normalizing it to $\sqrt{N}$ to account for the Poisson statistics of this counting experiment \citep{Wood_96a,Whitney_2011}.

For simplicity, in this paper we consider electron (Thomson) scattering only, both for the case of pure scattering (albedo $a=1$) and for the case of scattering plus hydrogen absorption ($a<1$). Although \textit{SLIP} has the capability to simulate polarized spectra, because electron scattering is a gray process, our results are monochromatic for the pure-scattering case. That is, these results are comparable to polarization observations at any wavelength. When we consider hydrogen absorption, we choose a representative optical wavelength of 6040 \AA~and discuss how absorption effects modify the pure-scattering results. At higher temperatures for which our calculated ionization fraction is very close to 1, these electron-scattering scenarios simulate a  fully ionized environment such as a region of shocked gas. This focus on electron scattering and single bow-shock structures distinguishes the simulations in this paper from those of \citealt{Shahzamanian_2016} and \citealt{zajacek_2017}. In Paper~II, we will present wavelength-dependent dust-scattering results from \textit{SLIP} and compare them with the bow-shock contribution to the polarization of the DSO as calculated by  \citealt{Shahzamanian_2016} and \citealt{zajacek_2017}.

\begin{figure}
 \includegraphics[width = \columnwidth]{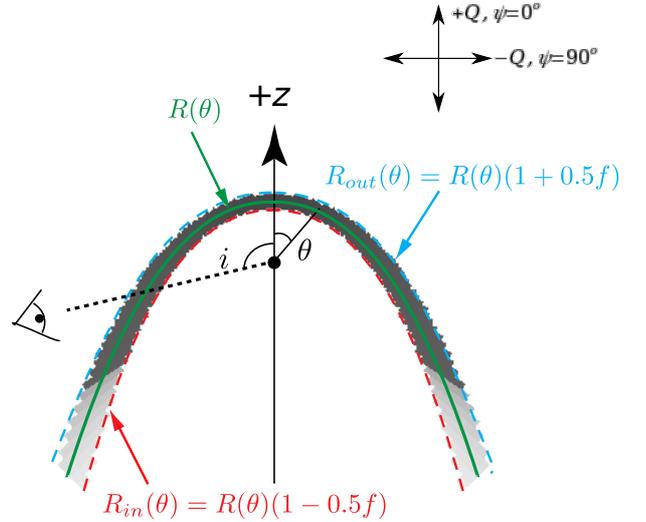}
\caption{Cross-section of our model geometry, along with a depiction of the bow shock density as a function of angle (greyscale).  The star is at the origin and moving in the direction of the arrow ($+z$). The central green solid line represents the central radius of the bow shock, which in our models we define with the Wilkin analytical solution  (Eq.~\ref{radwill}). Due to the difficulty of representing this equation graphically, in this figure we have used a graphical approximation of this function; however, the greyscale image is a discretisation of the actual Wilkin equation. The red and blue outer dashed lines represent our adopted inner and outer CSM radii, separated by a constant radial thickness $f$ as described in Section \ref{methods}. The density decreases from the bow head toward the wings of the shock (Eq.~\ref{rho}); we adopt an exponential decline in density in the far wings of the shock (Eq.~\ref{exp}).  The central source is shown exaggerated in size for reference. The angle $\theta$ is the polar angle measured from the $+z$ axis in our model grid, while the angle $i$ is the inclination or viewing angle for a distant observer.}
 \label{geom}
\end{figure}

Rather than simulating a particular object (as in \citealt{2013msao.confE.172N}, \citealt{Shahzamanian_2016}, and \citealt{zajacek_2017}), our goal here is to understand the polarization produced by electron scattering within a generalized bow shock. Thus, to describe our scattering region, we adopt the \cite{Wilkin_1996} analytic model of an axisymmetric bow shock formed when a star drives a wind into the stationary ISM while also moving along a straight line. This formulation assumes a spherically symmetric stellar wind and a locally uniform ISM.  The resulting bow shock structure and properties depend on the properties of the stellar wind, the speed of the star through the ISM, and the local ISM density. The solution provides for the shape, mass surface density, and velocity flow in an infinitesimally thin axisymmetric bow shock. The essential properties of this solution are the standoff radius of the bow head, the opening angle of the bow shock, and a characteristic surface density for the structure.

The standoff radius $R_0$ is defined as the location along the star's trajectory at which the ram pressures of the ISM and stellar wind are equal, i.e., $\rho_\text{w} V_\text{w}^2 = \rho_{I} V_\star^2$. Here $\rho_\text{w}$ represents the density of the stellar wind, $V_w$ the stellar wind velocity, $V_{\star}$ the stellar velocity, and $\rho_{I}$ the ISM density. With the stellar mass-loss rate represented by $\dot m_w$, this condition yields 

   \begin{equation}
R_0=\sqrt[]{\frac{\dot m_w V_w}{4\pi \rho_{I} V_\star^2}}
\label{standoff}
\end{equation}

\noindent  \citep{Wilkin_1996}.
Using momentum conservation and force balance, the bow shock radius as a function of polar angle is then given by 
  
 \begin{equation} 
   R(\theta)= \sqrt{3}R_0 \csc \theta \, \sqrt{1-\theta \cot\theta}\; .
   \label{radwill}
  \end{equation}
We use this equation to define the central radius of our model bow shock structure (Fig.~\ref{geom}). As described in \S~\ref{results}, we choose $R_0$ to give a convenient scale to our simulations. Note that at $\theta=\pi/2$, the extent of the bow shock is
 $\sqrt{3}R_0$.
 
\cite{Wilkin_1996} also determined the mass surface density $\sigma$ of the idealized, infinitesimally thin bow shock shell as a function of 
polar angle using conservation of momentum:

  \begin{equation}
    \sigma(\theta) = \frac{1}{2}\,R_0 \,\rho_{I} \frac{[2 \alpha (1-\cos\theta)+\tilde{\varpi}^{2}]^{2}}{\tilde{\varpi} \sqrt{(\theta-\sin\theta \cos\theta)^{2}+(\tilde{\varpi}^{2} - \sin^2\theta)^{2}}}\; .
    \label{sigma}
  \end{equation}
  
\noindent Here $\tilde{\varpi}$ is a convenient parametrization defined by $\tilde{\varpi}^{2} = 3(1-\theta \cot\theta)$. In the wings of the bow shock, $\tilde{\varpi} \gg 1$, giving $\sigma \propto \tilde{\varpi}$. The symbol $\alpha$ parametrizes the ratio of the translational speed of the star to its stellar wind velocity ($\alpha={V_*}/{V_w}$); in principle, the \cite{Wilkin_1996} model is valid only for $0<\alpha<1$.  When $\alpha = 0$, the stellar wind forms a spherical bubble and the standoff radius is undefined, whereas $\alpha > 1$ means the star is travelling faster than its wind.  For hot, massive stars with radiation-driven winds \citep{Cassinelli-textbook}, the wind velocity is much faster than that of the star, so that $\alpha \ll 1$.  On the other hand, for cool stars, the wind velocity can be slow relative to that of the star. For instance, the value of $\alpha$ for the O star $\zeta$ Pup is 0.1 \citep{Puls_1996}, while for Betelgeuse $\alpha$ is close to unity \citep{Mackey_Mohamed_Neilson_Langer_Meyer_2012}. In our models, we assume $\alpha=0.1$ to represent the hot-star case. 
%
%
 

Within \textit{SLIP}, it is not possible to encode an infinitesimally thin shell geometry with a divergent surface density.  Instead, we construct a finite scattering region that reproduces the mass surface density function from Equation \ref{sigma}. As noted above, we define the shock's mid-region with the Wilkin shape (Eq.~\ref{radwill}). Then we calculate the volume density necessary to match the Wilkin mass surface density (Eq.~\ref{sigma}) via $\rho (\theta) = \sigma(\theta) \,b(\theta)/\Delta R(\theta)$, where $\Delta R(\theta)$ is the radial thickness of the finite bow shock region. 
Here $b(\theta)$ is a geometrical correction factor arising from the $\theta$ dependence of the bow shock's radius; we discuss this factor in detail in Appendix \ref{appendix}.

Parametrizing the CSM thickness with the fractional quantity $f$ (where $f$ is constant over the shape and $0<f<1$), we calculate $\Delta R(\theta)$ as follows:
    
  \begin{equation}
  \Delta R(\theta)=R_{\textrm{out}}(\theta)-R_{\textrm{in}}(\theta) \equiv fR(\theta)\;.
  \label{chi}
  \end{equation}

\noindent In this equation, $R(\theta)$ is the radius of the bow shock at the interface of the ISM and stellar wind, given by Eq.~\ref{radwill}, $R_{\textrm{in}}(\theta)$ is the inner radius of the finite structure, and $R_{\textrm{out}}(\theta)$ is the outer radius. Approximations to these three functions are depicted as coloured lines in Fig.~\ref{geom}, while the actual discretized density is shown in greyscale. For a given value of $\theta$, $R_{\textrm{in}}$ and $R_{\textrm{out}}$ are equidistant from $R_0$.

We checked how changing the radial thickness $\Delta R(\theta)$ affects the simulated polarization signatures in the case of pure scattering ($a=1$). For values ranging from $f=0.1$ to $f=0.5$ (representing physically thin shells), we found insignificant variation in the polarization behaviour at any viewing angle.
Thus, in our simulations, we assume $f=0.25$, which ensures the thickness of the shell is at least one grid cell within the code structure.

With the definitions above, the volume density within our scattering region is given by

\begin{equation}
    \rho(\theta) = \frac{R_0 \rho_{I}b(\theta)}{2\Delta R(\theta)} \left\lbrace \frac{[2 \alpha (1-\cos\theta)+\tilde{\varpi}^{2}]^{2}}{\tilde{\varpi} \sqrt{(\theta-\sin\theta \cos\theta)^{2}+(\tilde{\varpi}^{2} - \sin^2\theta)^{2}}}\right\rbrace .
    \label{rho}
  \end{equation}

\begin{figure}
\includegraphics[width=\columnwidth]{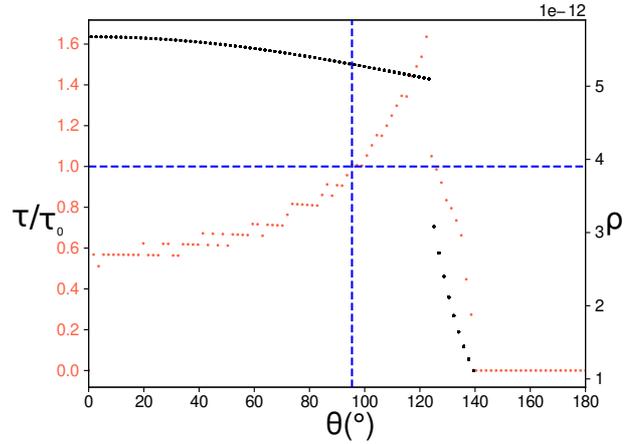}
\caption{Variation in mass density ($\rho$ [g cm$^{-3}$]; \textit{black points, right-hand axis}) and local normalized optical depth ($\tau/\tau_0$; \textit{red points; left-hand axis}) as a function of polar angle $\theta$. For each model, we specify the optical depth $\tau_0$ at the reference angle $\theta_0$ (\textit{dashed lines}; \S~\ref{methods}). The discrete nature of the optical depth is due to the distribution of the analytical bow shock shape across model grid cells. The behavior of the optical depth shows that the average number of scattering events per photon increases slowly with $\theta$ up to the cutoff angle (\S~\ref{results}) and decreases rapidly thereafter.}
\label{taudens}
\end{figure}

In the models presented here, we vary the density of the CSM by using as an input parameter the optical depth at a convenient arbitrary reference angle, $\theta_0=1.76 \textrm{ rad}=95.4^{\circ}$. We refer to this reference optical depth as $\tau_0$ and scale $\rho(\theta_0)$ to match it (effectively choosing $\rho_{I}$ to give the desired $\tau_0$). We then use Eq.~\ref{rho} to determine the density for other values of $\theta$. This results in a CSM density that is nearly, but not exactly, constant with $\theta$ (Fig.~\ref{taudens}). We then calculate $\tau(\theta)$ based on the density and thickness of the CSM. The variation of density and optical depth as a function of polar angle can be seen in Fig.~\ref{taudens}. The increase in optical depth with $\theta$ is due to the increasing behaviour of both $\sigma(\theta)$ (Eq.~\ref{sigma}; see discussion in \citealt{Wilkin_1996}) and $b(\theta)$ (Appendix~\ref{appendix}). 
To maintain a finite simulation size, we truncate the bow shock for large values of $\theta$ as described in Section \ref{results} below.

In the geometry of Fig.~\ref{geom}, $+Q$ Stokes vectors correspond to equatorial scattering, vertical polarization vectors (i.e., in the $\pm z$ direction), and polarization position angles near $\Psi=0\degr$. Negative or $-Q$ Stokes vectors correspond to polar scattering, horizontal polarization vectors (i.e., in the plane orthogonal to $\pm z$), and position angles near $\Psi=90\degr$. Stokes $\pm U$ denotes diagonal polarization vectors rotated $45\degr$ from the $\pm Q$ vectors. (In our axisymmetric models, $U$ averages to zero for unresolved cases.) Because we consider only electron scattering, a symmetric process, our models produce no Stokes $V$ (circular) polarization. Thus, the fractional polarization $p$ (usually expressed as a percentage) is defined as
\begin{equation}
p(\%) = \frac{\sqrt[]{Q^2+U^2}}{I} \times 100.
\end{equation}

\section{Results from analytical model}\label{analytic}

Before embarking on a parameter study using the MCRT methods of the \textit{SLIP} code, we first consider semi-analytic results for scattering within a bow shock in the optically thin limit. Because the stellar wind bow shock of \cite{Wilkin_1996} is explicitly axisymmetric, the methods of \cite{brown_1977} can be used to determine its expected polarization as a function of viewing angle in the spatially unresolved case.

\cite{brown_1977} derived a simple expression for the linear polarization from an axisymmetric and optically thin scattering region illuminated by a central point source.  Considering scattered light only, 
the fractional polarization can be expressed as

\begin{equation}
p=\frac{\sin^2  i}{h(\gamma)+\sin^2 i}\;,
\label{pdef}
\end{equation}
\noindent where $i$ is the viewing angle measured from the $z$-axis as shown in Fig.~\ref{geom}, $\gamma$ is a ``shape factor'' to be discussed below, and $h(\gamma) = 2(1+\gamma)/(1-3\gamma)$. Brown \& McLean use the symbol $\alpha$ in the expression for $p$ (their Eqn.~17), but we choose to define $h(\gamma)\equiv2\alpha$ because we have already introduced a different $\alpha$ in the context of the bow shock geometry.

The shape factor $\gamma$ is given by

\begin{equation}
\gamma =\frac {\int_{r=0}^{\infty}\int_{\mu = -1}^{1} n(r,\mu) \mu^2 dr d\mu}{\int_{r=0}^{\infty}\int_{\mu = -1}^{1} n(r,\mu) dr d\mu,}
\label{gamma}
\end{equation}

\noindent where $\mu =\cos{\theta}$ (with $\theta$ representing the polar angle measured from the $z$-axis; Fig.~\ref{geom}) and $n(r,\mu)$ is the number density of the scattering region \citep{brown_1977}. Values of $\gamma$ range from 0 to 1, with $\gamma=1/3$ representing a spherical envelope, $\gamma=0$ a planar disk, and $\gamma=1$ a bipolar jet. These geometries produce maximum polarization values (at viewing angles of $90\degr$) of $0\%$, $33\%$, and $100\%$ respectively. In the specific case of the Wilkin model, we have

\begin{equation}
n(r,\mu) = \frac{\sigma(\mu)  }{\Delta R(\mu)}\; .
\label{column}
\end{equation}

\noindent When we substitute our expressions for $\sigma$ from Eq.~(\ref{sigma}) and $\Delta R$ from Eq.~(\ref{chi}) into Eq.~\ref{column}, and then put the resulting expression for $n(r, \mu)$ into Eq.~\ref{gamma}, we determine the shape factor $\gamma$  for our modified Wilkin bow shock. Because the bow shock is not a closed shape, we take the angular integrals from $\theta=0\degr$ to $\theta=131\degr$ only. The resulting $\gamma$ factor depends only on $f$, the fractional thickness of the shell, and $\alpha$, the velocity ratio (both defined in Section \ref{methods}). Numerical evaluation of the integrals in Eq.~\ref{gamma} for $f=0.25$ and values of $\alpha$ between 0.1 and 10 yields $\gamma \approx 0.241-0.295$. Corresponding values of $h(\gamma)$ range from $8.96$ to $22.52$. 

Given these generally large values of $h(\gamma)$, we expect that for low scattering optical depths, the polarization should scale with viewing inclination as $p \propto \sin^2 i$, which is symmetric about $i=90^\circ$. For representative values of $\alpha = 0.1$ and  $h(\gamma) = 8.96$, we conclude that the theoretical electron-scattering polarization for an unresolved bow shock structure is

 \begin{equation} 
   p(\%)=11.16 \, \sin^2 i .
   \label{analyteq}
  \end{equation}

\begin{figure}
\includegraphics[width=\columnwidth]{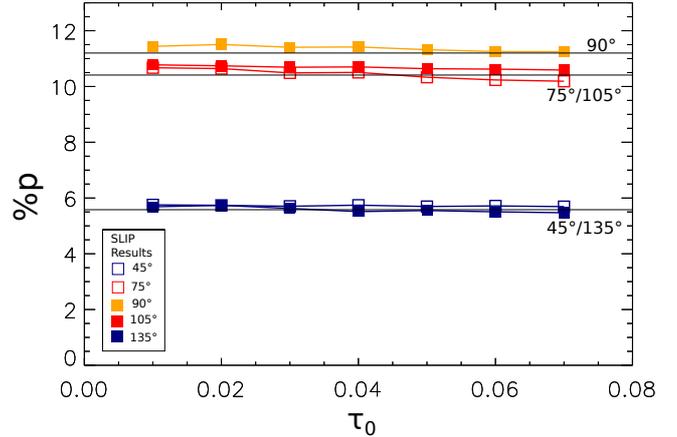}
\caption{Fractional polarization (with respect to scattered light only) as a function of optical depth at the standoff radius ($\tau_0$) for \textit{SLIP} models of an optically thin, unresolved bow shock viewed at $i=90^{\circ}$ (\textit{gold}), $i=75^{\circ}$ and $105^{\circ}$ (\textit{red}), and $i=45^{\circ}$ and $135^{\circ}$ (\textit{blue}). Horizontal lines represent the analytical prediction (symmetric about $i=90^{\circ}$) for each angle (Eq.~\ref{analyteq}). Our numerical simulations reproduce the theoretical predictions well, with some expected deviation from symmetry at larger optical depths. Error bars representing $1\sigma$ uncertainties in each model bin (\S~\ref{methods}) are smaller than the plotted symbols.}
\label{pvopt_pflux}
\end{figure}

We constructed a set of \textit{SLIP} models with $f=0.25$, $\alpha=0.1$, and $a=1$, with photons arising from the central source only, to compare with these analytical results (Fig.~\ref{pvopt_pflux}). We considered reference optical depths of $\tau_0\leq 0.07$ only in order to ensure that the average number of scatters per photon was very close to 1. Our simulations show a viewing angle dependence and symmetric behaviour about $90^{\circ}$ in agreement with the prediction of Eq.~\ref{analyteq}, which serves to verify that our numerical approach is valid. The values arising from the simulation are generally consistent with the analytic model for these optical depths, with small differences attributable to our discretization of the Wilkin function for the \textit{SLIP} models. The symmetry about $90^{\circ}$ begins to break down slightly as $\tau_0$ increases, which is expected given the variation in actual optical depth with viewing angle (Fig~\ref{taudens}).



\section{Model predictions from \textit{SLIP}}
\label{results}

In order to perform numerical calculations of the polarization created in a Wilkin bow shock, we must take into account the fact that our simulations involve a grid of finite size, whose maximum extent we set at $R_{\textrm{max}} = 6.68$ AU.  Our approach is to modify the density description in the \cite{Wilkin_1996} model to accommodate our finite grid.  We use the density of the bow shock as prescribed by Eq.~\ref{rho}, up to a certain cutoff angle $\theta_c$. For $\theta>\theta_c$, we assume the bow shock density declines exponentially rather than being sharply truncated by the outer limit of our simulation (which we found resulted in spurious polarization at the edges).  This modified density in the wings of the bow shock is given functionally by

 \begin{equation}
    \rho(\theta > \theta_{\rm c}) = \rho(\theta_c,\varpi) \exp[-(\theta-\theta_c)/\delta\theta_0]\; ,
    \label{exp}
\end{equation}

\noindent where $\delta\theta_0$ is a constant angle governing the steepness of the density decline. 

This modification of the Wilkin density structure does not affect the accuracy of our results, for two reasons. First, an infinitesimally physical thin shell is not  physically realistic, especially at large distances from the bow head, as the shell must spatially ``thicken'' with distance by virtue of gas pressure gradients and Kelvin-Helmhotz instabilities \citep{Mohamed_2012,Mackey_Mohamed_Gvaramadze_Kotak_Langer_Meyer_Moriya_Neilson_2014}. Second, the geometry for a thin shell ensures that with increasing distance from the star, the solid angle subtended by a shell ring (i.e., a ring about the symmetry axis) decreases with distance.  As a consequence, from the perspective of scattering stellar photons, the large-scale wings of the bow shock offer a diminishing cross-section for intercepting and scattering starlight. This also means that the increasing size of the grid cells at larger radii does not significantly affect our results.

We investigated the impact on polarization of the cutoff angle $\theta_c$ and the steepness $\delta\theta_0$ of the exponential decay function by varying both parameters in our simulations. We emphasize that in these and all our subsequent models we measure fractional polarization with respect to the total light, rather than scattered light only (as in Eqns. \ref{pdef} and \ref{analyteq}). 

In testing the effects of $\theta_c$ and $\delta\theta_0$, we used the central photon source with reference optical depth of $\tau_0=0.5$ and a CSM temperature $T$ of 10,000 K. For an unresolved bow shock, we found that as the cutoff angle increases, the peak polarization value and the variation of polarization with viewing angle $i$ is nearly unchanged. We thus chose a convenient value of $\theta_c=2.1$ rad ($122^{\circ}$) as the cutoff angle for all the other models presented in this paper. This choice for $\theta_c$ ensures that the entire CSM structure is included within our simulation grid. 
All the values we tested for $\delta\theta_0$ resulted in similar polarization values and behaviour. We chose $\delta\theta_0=0.3$ rad ($17^{\circ}$) for all the models shown hereafter.

We also tested the behaviour of the polarization in our simulations as a function of $\alpha$, the velocity ratio defined in Section~\ref{methods}. Fixing the albedo of the scattering region at $a=1$, emitting photons from the central source, and using the same values of $\tau_0$ and $T$ as in our previous test cases, we found that as $\alpha$ increases, the polarization value increases as well. From Equation~\ref{rho}, we see that with a given thickness function $\Delta R(\theta)$, the volume density $\rho$ increases with $\alpha$ for angles greater than $\theta=0$. Thus, increasing the value of $\alpha$ should have a similar effect to increasing the optical depth $\tau_0$ for $a=1$, which does indeed increase polarization overall (Section~\ref{thomson}). For the simulations presented below, we set $\alpha=0.1$ as discussed in Section~\ref{methods}.

Finally, we studied how changing the standoff radius $R_0$ of the bow shock changes the polarization behaviour. When the albedo $a$ is fixed at 1 (the pure scattering case), changing $R_0$ does not affect the polarization. 

However, when the albedo is not explicitly fixed  (the case of scattering with absorption), changing the standoff radius changes the albedo and thus the polarization. This is because $R_0$ is used to calculate the physical thickness $\Delta R(\theta)$ of the bow shock (Eqs.~\ref{radwill} and \ref{chi}), which in turn affects its opacity. When $a$ is not fixed, it is calculated using the opacity of the region (\S~\ref{vara}): a larger value of $R_0$ corresponds to a lower density for a given $\tau_0$, which leads to a larger opacity and a lower albedo.

We chose $R_0 = 1.4$ AU for all our models, because for variable $a$ this $R_0$ value produces polarization behaviour as a function of viewing angle similar to the analytical results in the optically thin case (Section~\ref{analytic}). (For comparison, the radius of our central source is $1 R_\odot\approx0.005$ AU; this value has no physical significance other than to make the central star effectively a point source.) With $R_0 = 1.4$ AU and $R_{\textrm{max}} = 6.68$ AU, the density within the bow shock goes to zero between $\theta=134\degr$ and $\theta=140\degr$ (where the bow shock radii intersect the boundary of the simulation). 


To create our numerical simulations, we used the University of Denver's high-performance computing cluster (HPC), which consists of 180 Intel Xeon processors running at 2.44 GHz. Each of our model runs used 16 CPUs with $10^8$ photons per CPU. This yielded polarization uncertainties on the order of $\sigma_p(\%) \sim 0.01$. Completing each run took $\sim 60-70$ minutes, with slightly longer times for larger values of $\tau_0$.  
Our simulations can be broadly divided into models assuming pure Thomson scattering with no absorption ($a=1$) and those including some absorption (variable $a$). In each case, we studied the effect of various parameters on the polarization behaviour for both resolved and unresolved cases. In the resolved cases, we preserve spatial information from our simulations, while in the unresolved cases, we combine all photons from a given viewing angle into a single set of polarization values. We present our results below. 

\subsection{Pure Thomson Scattering}
\label{thomson}
To simulate the case of pure Thomson scattering, we fixed the albedo of the bow shock environment at 1. In this case, all emitted photons scatter in the bow shock and ultimately escape. We explored the dependence of polarization on CSM temperature, standoff radius, and optical depth for both central and distributed photon sources. We found that for a given source, only the optical depth affects the simulated polarization; varying the CSM temperature and standoff radius produced no change in either polarization magnitude or behaviour as a function of viewing angle. 

In the rest of this section, we present the detailed behaviour of polarization as a function of optical depth, for both resolved and unresolved scenarios. We investigated three representative optical depths: $\tau_0=0.1, 0.5$, and $2.0$. In all the cases shown here, $T=10,000$ K, $\theta_c = 122^{\circ}$, $\delta\theta_0 = 17^{\circ}$, and $\alpha = 0.1$. In all these simulations, we found polarization position angles very close to $\Psi=0\degr$, so we have not displayed the position angle results.

\subsubsection{Optical depth dependence -- resolved bow shock}
\label{result_opt_r}

\begin{figure*}
\includegraphics[scale = 0.8]{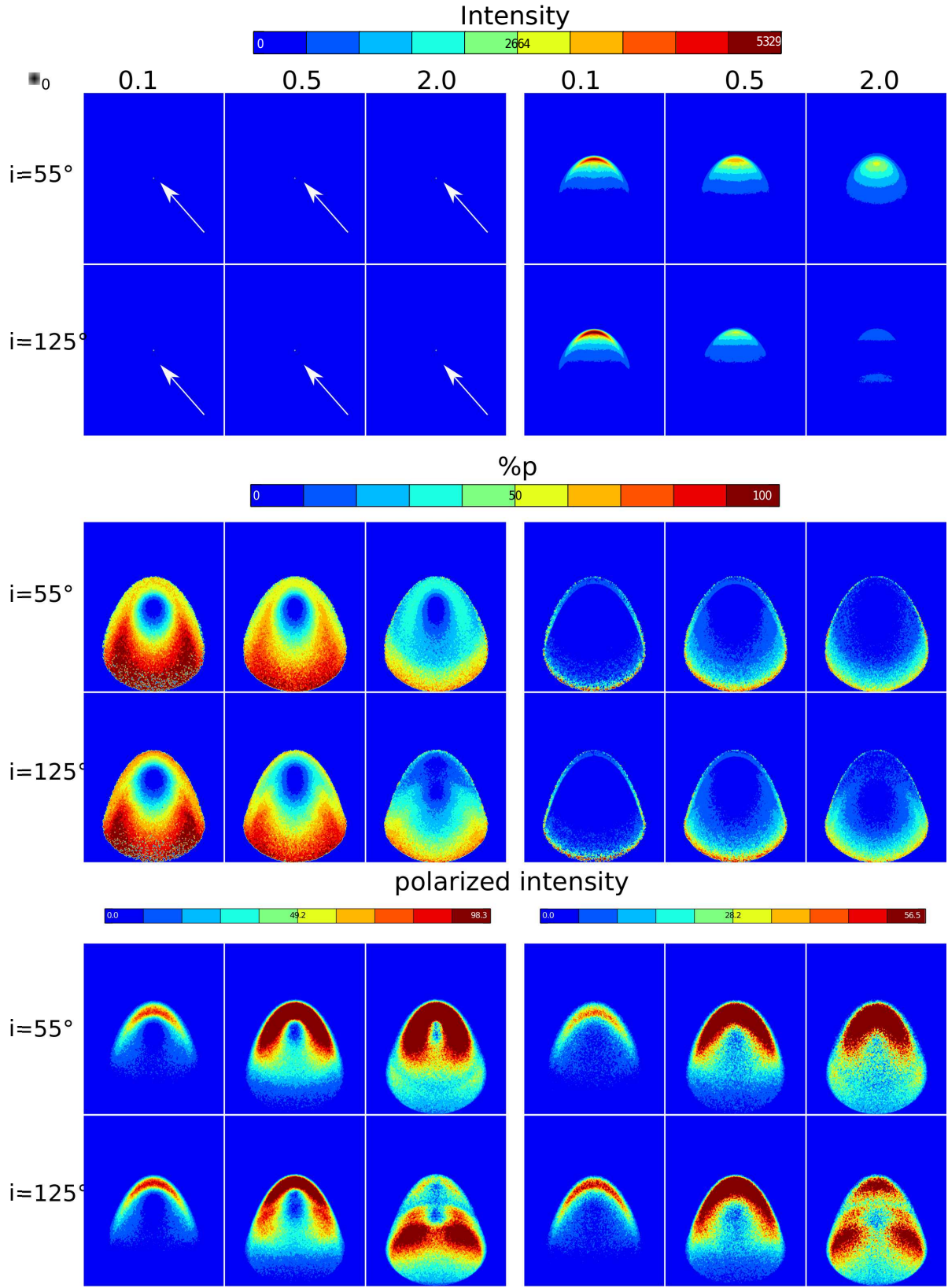} 
\caption{Intensity, polarization, and polarized intensity maps for resolved bow shocks illuminated by a central source (\textit{left}) and the distributed source (\textit{right}; photons arise from within the CSM as described in \S~\ref{methods}). In the central-source intensity maps, arrows indicate the location of the star.  We show two inclination angles symmetric about $90^{\circ}$. Optical depth increases from left to right in each row. Intensities are in arbitrary units. }
\label{map_diffopt}
\end{figure*}

In Fig.~\ref{map_diffopt}, we display the intensity, percent polarization, and polarized intensity images for a resolved bow shock with three different optical depths at two representative inclination angles symmetric around the $z=0$ plane, $55^{\circ}$ and $125^{\circ}$.~(Polarized intensity is calculated by multiplying $\%p$ by intensity; in these maps it represents the polarized light arising from the system.) In the central-source cases (left column), the intensity maps show only a small dot at the location of the star due to our choice of a linear intensity scale that shows the distributed-source behavior well. The scattered light from the bow shock contributes intensity too faint to be seen on this scale.

\begin{figure*}
\includegraphics[scale=0.7]{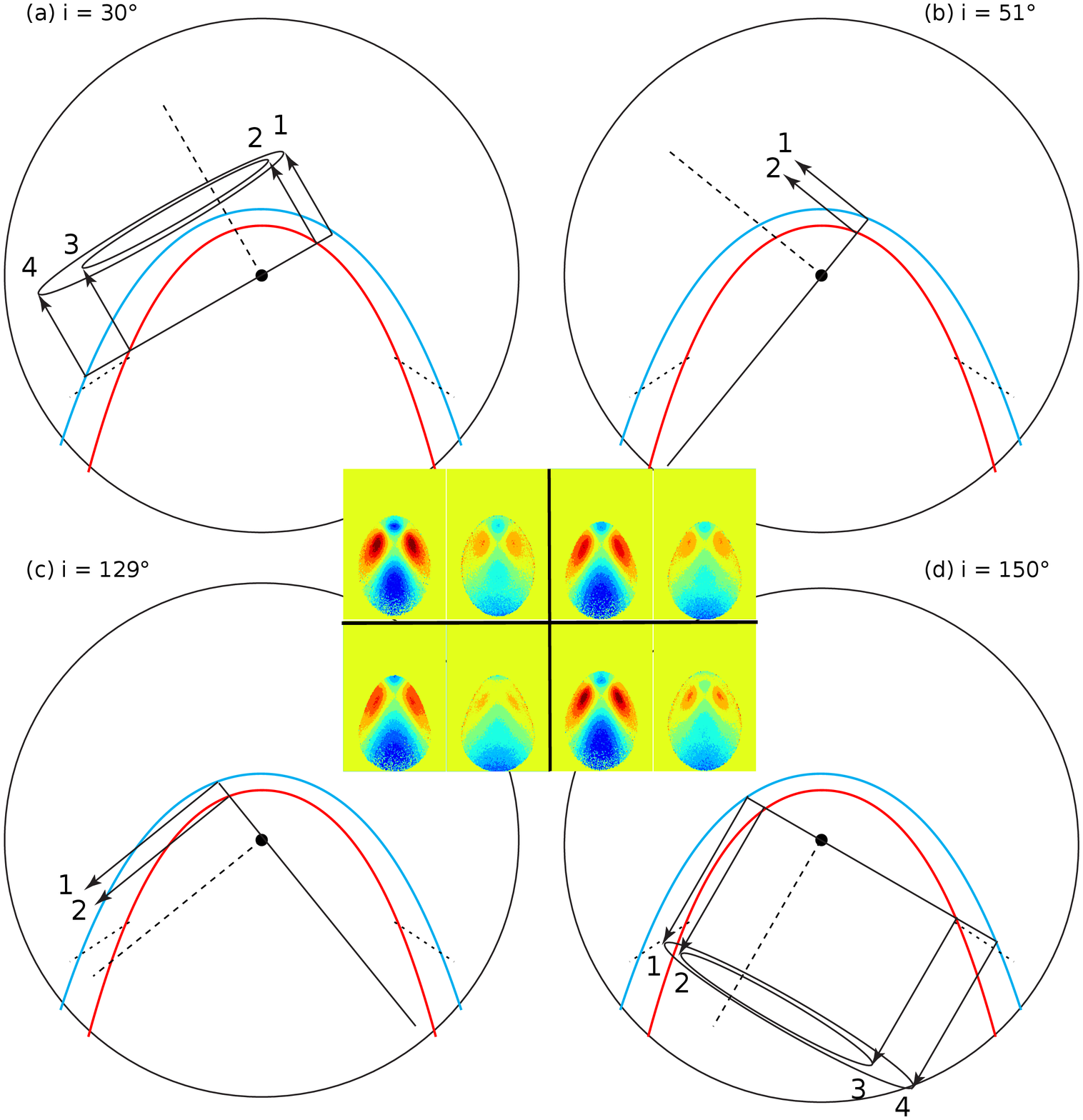} 
\caption{Sketch showing the $90^{\circ}$ scattering paths for central-source photons at four different viewing angles $i$. The numbered arrows indicate the limiting paths that produce negative $q$ polarization as seen by an observer in the $i$ direction (polar scattering). In each panel, there will also be  $90^{\circ}$ scattering paths for photons initially directed out of the page, defining the width of the scattering ellipses; these paths, which produce positive $q$ polarization (equatorial scattering), are not shown in the sketch. Dashed lines indicate the direction to the observer; short dotted segments mark the location of the density falloff in the wings of the bow shock (Section~\ref{results}). Small coloured images for each inclination angle depict the distribution of $q$ polarization as seen by the observer, for $\tau_0 = 0.1$ (\textit{left}) and $\tau_0 = 2.0$ (\textit{right}). The colours range from $-100\%$ (\textit{darkest blue}) to $+100\%$ (\textit{darkest red}).}
\label{bowsketch_los}
\end{figure*}

The central-source polarization maps are similar for the two symmetric inclination angles; they show a generally elliptical polarization pattern, which is created by the combination of all $90^\circ$ scattering paths, as shown schematically in Fig.~\ref{bowsketch_los}. For a given inclination angle, the overall polarization magnitude decreases with increasing optical depth, which is generally expected given that multiple scatters typically randomize the polarization of an ensemble of photons. For a given optical depth, the polarization near the bow head is smaller for the larger inclination angle. Figure~\ref{bowsketch_los} shows that the path length for photons scattering at $90^\circ$ near the bow head at the lower inclination angle (panel $b$, paths 1 and 2) is much smaller than in the case of the higher inclination angle (panel $c$, paths 1 and 2). Because of this, multiple scattering is more important for higher inclinations and optical depths. In this case, because the outgoing photons scatter in the same plane, the dominant effect of multiple scattering is to remove polarized photons from the beam rather than randomizing their position angles. This effect can be seen in the decrease of polarized intensity with inclination angle in the lower panels (Fig.~\ref{map_diffopt}).

The central-source polarized intensity maps show that the majority of scattered photons reach our line of sight from locations near the bow head; the scattering material is very tenuous in the outer regions, so very few photons scatter there (but those that do become highly polarized in the process). We note that although the resolved maps look similar in polarization between the two angles, they are quite distinct in polarized intensity, particularly at higher optical depths. This suggests that polarized intensity maps may provide an observational tool for constraining bow shock inclinations.

In the distributed-source case (photons arising only from the CSM; \textit{right side}), Fig.~\ref{map_diffopt} shows that the total intensity is concentrated near the bow head because the CSM density is higher in that region and thus more photons are emitted from there. In this case, photons are emitted with an isotropic distribution of initial directions from within the volume of the CSM. Thus, photons scatter more times on average than in the central-source model with the same input parameters. This increased scattering, combined with cancellation from neighbouring photon origins and the contribution from ``surface" photons (those arising from the outer edge of the bow shock) that reach the observer directly, causes a significant decrease in the polarization arising from any given location in the CSM, compared with the central-source case (middle panels of Fig.~\ref{map_diffopt}). The polarization is highest at the edge of the CSM because of a scattering asymmetry. In most parts of the CSM, polarization angles are highly randomized, so photons that reach the viewer can have any polarization angle. However, limb photons cannot scatter in all directions and thus tend to have a preferred polarization angle. The difference in polarization morphology between central-source and distributed-source models suggests that observational polarization maps \citep[such as those of][]{2013A&A...551A..35R} can be useful for constraining the photon origin and thus the relative brightnesses of the star and the CSM.

By contrast, the distributed-source polarized intensity maps look very similar to those produced by the central-source models and show similar variations with inclination and optical depth. Thus, observed polarized intensity maps would not be able to distinguish reliably between photons emitted from the central star and photons emitted from the bow shock.

\subsubsection{Optical depth dependence -- unresolved bow shock} 
\label{result_opt_ur}

\begin{figure*}
\includegraphics[width=\textwidth]{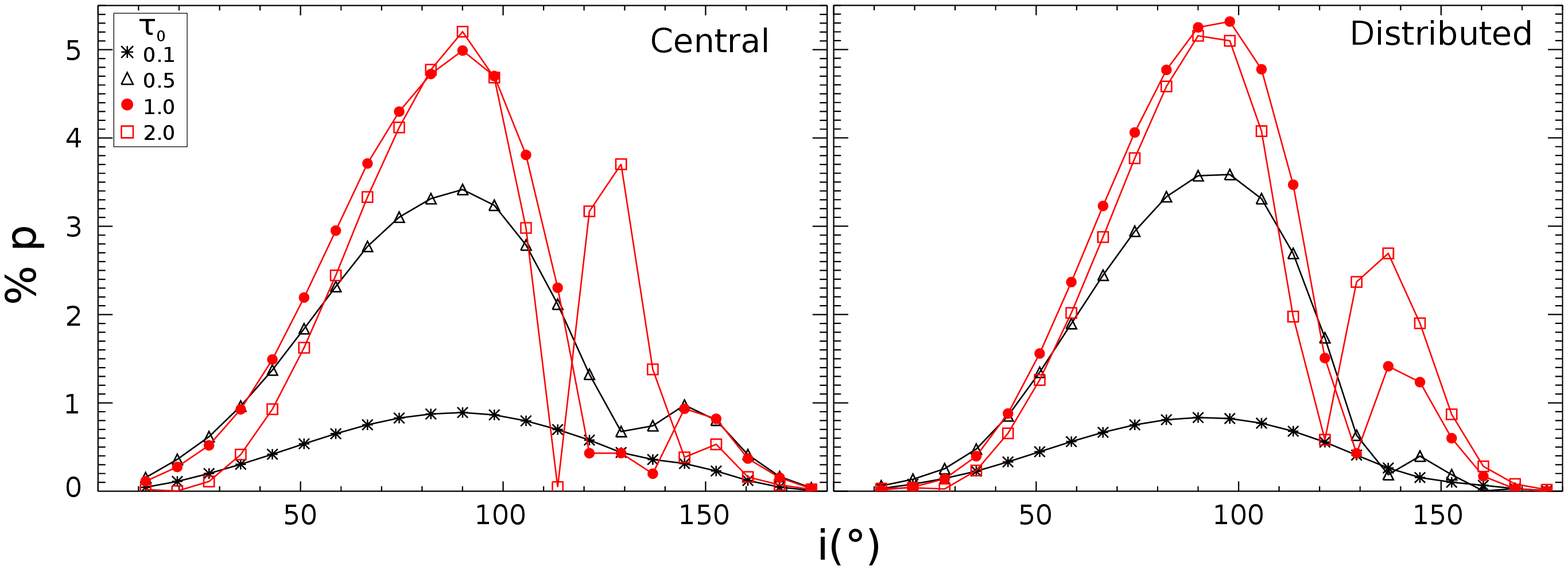} 
 \caption{Polarization as a function of inclination angle for an unresolved bow shock with different values of $\tau_0$, for photons arising from the central source (\textit{left}) and from the CSM (distributed-source; \textit{right}). All other parameters are held constant as described in \S~\ref{thomson}. Error bars representing $1\sigma$ uncertainties in each model bin (\S~\ref{methods}) are smaller than the plotted symbols.}
 \label{pva_opt}
\end{figure*}

In Fig.~\ref{pva_opt}, we display the polarization variation as a function of viewing angle for the unresolved case, considering four different values of the reference optical depth $\tau_0$. For both central and distributed emission cases, all models show a primary peak in percent polarization at an inclination angle of $90\degr$, as well as a secondary peak at angles greater than $130\degr$ whose exact location depends on $\tau_0$.  

In Fig.~\ref{pva_opt}, the maximum polarization occurs at an inclination angle of $90^{\circ}$ for all optical depths and both photon sources. This can be understood in terms of the analytical models of \citet{brown_1977}, who showed that for the optically thin case, the polarization produced by scattering in an axisymmetric envelope is proportional to $\sin^2 i$.

\begin{figure*}
\includegraphics[width=\textwidth]{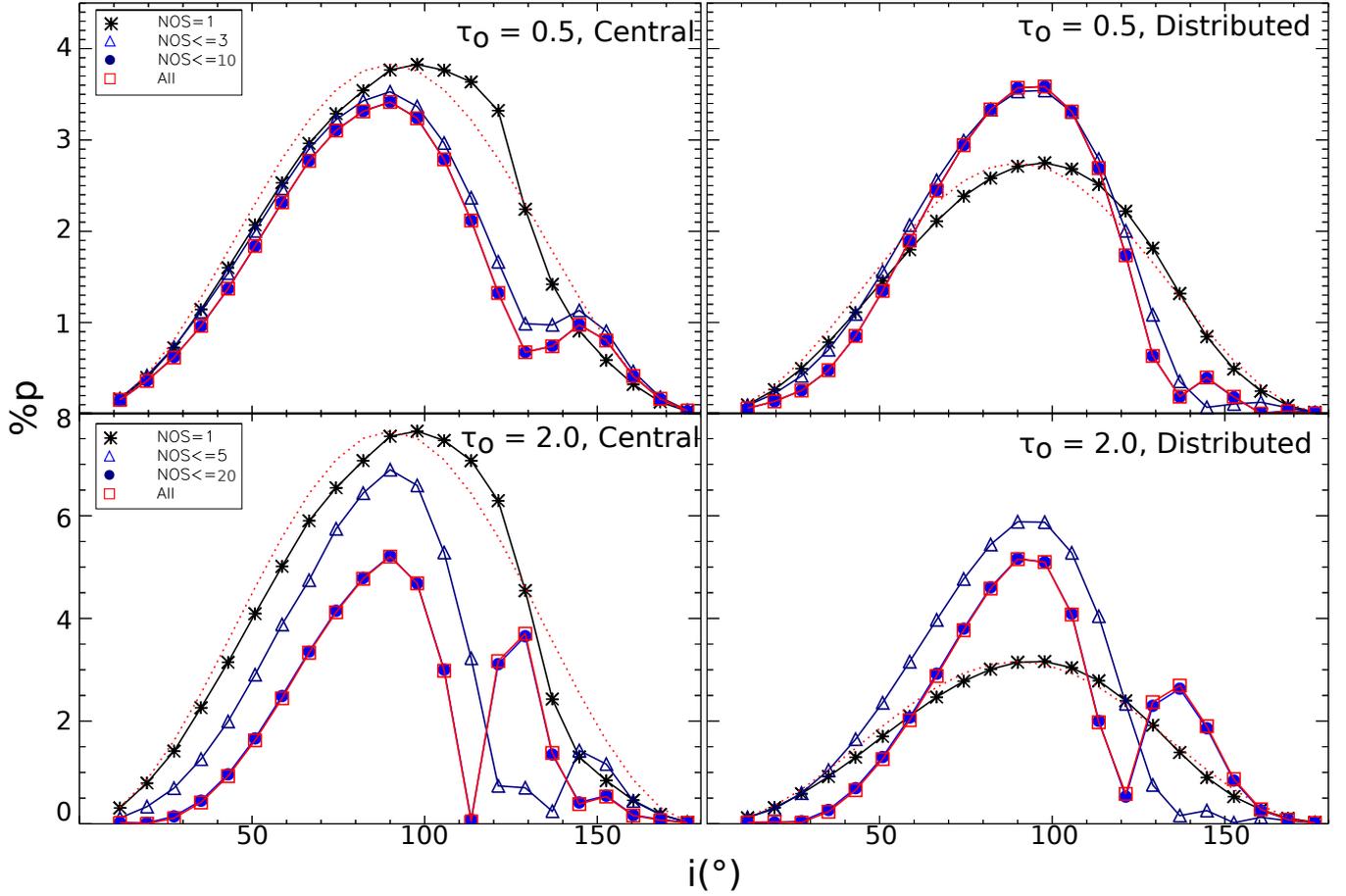} 
\caption{Polarization as a function of inclination angle for the same models as in Fig.~\ref{pva_opt}, with different curves for photons scattered different numbers of times. In the legends, ``NOS" denotes number of scatters. ``All" refers to the photons that have been scattered any number of times. The red dotted line in each panel traces the theoretical $\sin^2 i$ function \citep{brown_1977}, normalized to the peak of the single-scattering curve in each panel. Error bars representing $1\sigma$ uncertainties in each model bin (\S~\ref{methods}) are smaller than the plotted symbols.}
\label{pva_le}
\end{figure*}

For higher $\tau_0$ values, however, our models depart from the theoretical $\sin^2 i$ dependence of the polarization, particularly at higher viewing angles. As the optical depth increases, the secondary peak becomes enhanced with respect to the primary peak, and even exceeds it at larger optical depths than we display here. (We tested a range of $\tau_0$ values to establish this behaviour, but only display a few in Fig.~\ref{pva_opt} for clarity.) We hypothesize that this effect is due to multiple scattering becoming more common at higher optical depths. In order to understand the effect of multiple scattering on the polarization behaviour, we created central-source and distributed-source simulations for  $\tau_0 = 0.5$ and  $\tau_0 = 2.0$ in which we disaggregated the results by number of scatters; we display the results in Fig.~\ref{pva_le}. Indeed, we see from this figure  that the singly scattered photons is consistent with the theoretical $\sin^2 i$ dependence (with a slight ``shoulder" at low $\tau_0$ due to the onset of the density falloff; Eq.~\ref{exp}. Other slight departures from the idealized function are due to the discretisation effects discussed in \S~\ref{analytic}). The multiply scattered photons diverge from this behaviour more strongly as $\tau_0$ increases, particularly at larger viewing angles where the path length through the CSM is longer (Fig.~\ref{bowsketch_los}). 

\begin{figure*}
\includegraphics[width=\textwidth]{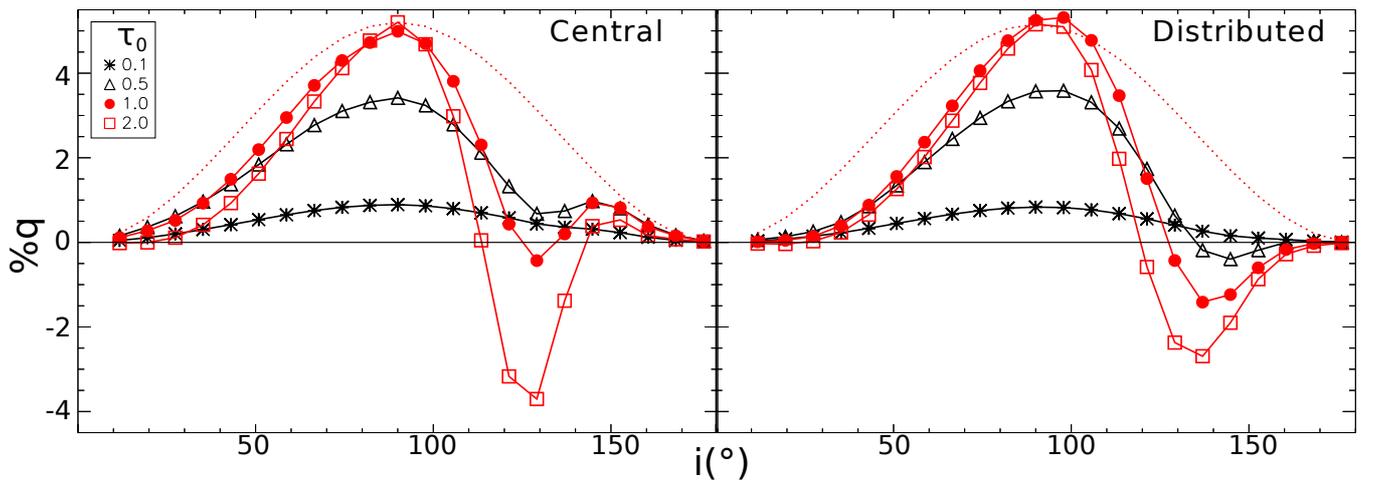}
 \caption{Percent Stokes $q$ polarization as a function of inclination angle for four different values of optical depth $\tau_0$, for photons arising from the central source (\textit{left}) and from the CSM (distributed source; \textit{right}). Black points and lines represent optically thin cases, while red points and lines represent higher optical depths. Red dotted lines represent the theoretical $\sin^2(i)$ function normalized to the peak of the $\tau_0 = 2.0$ curves. Error bars representing $1\sigma$ uncertainties in each model bin (\S~\ref{methods}) are smaller than the plotted symbols. Positive values of $q$ correspond to polarization position angles of $\Psi=0\degr$, while negative values correspond to $\Psi=90\degr$.}
 \label{qva_opt}
\end{figure*}

We also see that the overall width of the polarization curve decreases for larger numbers of scatters (Fig.~\ref{pva_le}), particularly at higher optical depths. We attribute this to the increasing contribution from scattering paths producing negative $q$ (``polar scattering") polarization in these cases. (Stokes $u$ is zero on average for these axisymmetric models, so $q$ is the dominant contributor to the total polarization $p$.) In the central-source case, the scattering paths producing positive $q$ polarization (``equatorial scattering") have a constant average initial (pre-scattering) path length through the CSM independent of viewing angle; thus the $+q$ polarization varies as $\sin^2i$ due to projection effects. (These positive-$q$ paths are not shown in Fig.~\ref{bowsketch_los}: they initially run from the central source directly out of the page, then scatter toward the observer in the direction indicated by the arrows. They create the red regions in the inset $q$ maps.) By contrast, the negative-$q$ paths shown in Fig.~\ref{bowsketch_los} have path lengths through the CSM that vary with inclination angle, and these are longer than the $+q$ paths for most angles. This means that increasing optical depth results in a higher magnitude of $-q$ polarization, as shown explicitly in Fig.~\ref{qva_opt}. With no absorption, more photons scatter into other lines of sight, while the few that escape toward the observer have scattered multiple times in the same plane and are thus more highly polarized \citep[as discussed in][]{Wood_96a}. On the other hand, higher optical depths and more scatters produce more negative $q$ polarization and smaller values of $p$ in Fig.~\ref{pva_le}. For the viewing angles with negative $q$ values, the polarization position angle $\Psi$ flips from $0\degr$to $90\degr$.

We therefore conclude that the secondary peak near $i=130\degr$ in the unresolved, central-source models with higher optical depths (Fig.~\ref{pva_opt}) is caused by a strong increase in $-q$ polarization when multiple scattering becomes important. Most of our models also show a polarization peak near $150\degr$ due to the fact that at this angle, the line of sight no longer intersects the near side of the CSM because of our simulation boundary (\S~\ref{results}). In this case, the path lengths that pass through the near side of the CSM are very long, so almost no photons escape there; the resulting polarization is primarily due to photons that are singly scattered from the interior far wall of the CSM (path 3 in Fig.~\ref{bowsketch_los}, panel $d$).

In the distributed case, the polarization predominantly arises from the limb of the bow shock and from the wings farthest from the bow head (Fig.~\ref{map_diffopt}). Photons from the limb tend to produce $+q$ polarization (in addition to some $u$, which cancels out in the unresolved case) because they are most likely to reach the observer by singly scattering near the edge of the CSM, producing the familiar tangential polarization pattern. Photons arising from the plane facing the observer produce zero net polarization because they are equally likely to escape after scattering in any direction, and thus cancellation is high. In the wings, however, this symmetry breaks due to the density falloff; in this case photons are most likely to escape after singly scattering in the regions farthest from the bow head, producing negative $q$ values.

For the unresolved distributed models (Fig.~\ref{pva_opt}), the polarization as a function of viewing angle behaves very similarly to the case of the central-source models, as expected because the bow-shock geometry of the CSM is the same between the two cases \citep{brown_1977}. We see the same $\sin^2 i$ behaviour, modified by increasing contributions from $-q$ polarization at higher viewing angles (Fig.~\ref{qva_opt}) as we see more contribution from the far side of the bow shock. The secondary peak in the distributed case occurs at larger viewing angles than in the central-source case because the CSM density falloff translates into fewer photons emitted from those angles.

Interestingly, although the central-source and distributed models show very similar polarization behaviour as a function of optical depth (Fig.~\ref{pva_opt}), they behave quite differently as a function of number of scatters for a given optical depth (Fig.~\ref{pva_le}). In the distributed models, multiple scattering \textit{increases} the polarization over single scattering at intermediate viewing angles. We attribute this effect to the fact that polarization in the distributed cases arises primarily from the limb, where column densities are high. Although this polarization is likely dominated by singly scattered photons originating near the outer surface, a few multiply scattered photons reaching us through the dense material at the limb can create large polarization percentages due to scattering in the same plane \citep{Wood_96a}. For higher optical depths and more scatterings, however, the two emission cases become quite similar, as expected once the photon source becomes ``forgotten.''

\begin{figure*}
\includegraphics[width=\textwidth]{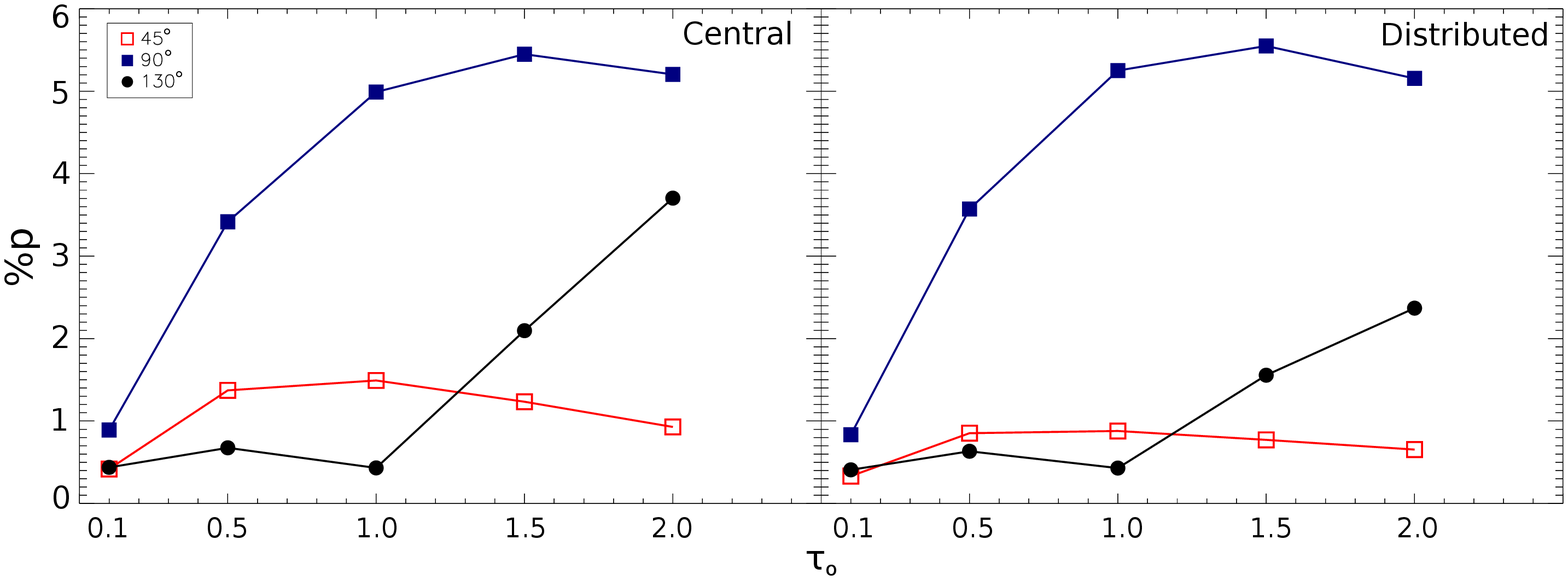}
\caption{Polarization as a function of optical depth $\tau_0$ at three different inclination angles (labelled in degrees), for photons arising from the central source (\textit{left}) and from the CSM (distributed source; \textit{right}). Error bars representing $1\sigma$ uncertainties in each model bin (\S~\ref{methods}) are smaller than the plotted symbols.}
 \label{pvtau_opt}
\end{figure*}

In Fig.~\ref{pvtau_opt}, we compare the variation of polarization with optical depth for three different inclination angles and the two photon sources. As expected based on previous results, the central-source and distributed cases show similar behaviour. For the lower viewing angles, we see the ``peaking'' effect described by \citet{Wood_96a}, in which polar scattering begins to dominate over equatorial scattering for higher optical depths.
At $i=45^\circ$, the polarization magnitude is relatively low for all $\tau_0$ values due to large contributions from $-q$ scattering paths (Fig.~\ref{bowsketch_los}). At $i=90^\circ$, the location of the first polarization peak in all our models, the polarization is a maximum for all optical depths due to the loss of paths 3 and 4 combined with a very short path length through the CSM at the bow head for paths 1 and 2 (which allows more photons to escape without scattering).
At  $i=130^\circ$, the location of the second polarization peak for the central-source case, the behaviour is quite different: our models show a dramatic \textit{increase} in polarization magnitude as a function of optical depth for $\tau_0>1$, with central-source models increasing more steeply than distributed models. At this inclination angle, the path lengths for scattering producing $-q$ polarization are at their longest (Fig.~\ref{bowsketch_los}$c$); increasing optical depth increases the number of scatterings photons undergo in the same plane, while filtering out photons with lower polarization; this increases the $-q$ contribution as discussed above. Hence, the polarization increases with increasing optical depth, and the effect is more pronounced for the central-source models because the path lengths through the CSM are longer in these cases.

Our results can be used along with observational data to constrain the inclination angle and optical depth of a given bow shock nebula, assuming electron scattering is the primary polarizing mechanism. An unresolved bow shock would be observed at a single value of $i$ and $\tau_0$. Once corrected for interstellar polarization (and for orientation on the sky in the case of $q$, e.g. via proper motion measurements), observed values of $p$ and $q$ for such an object would yield horizontal lines in Figs.~\ref{pva_opt}, \ref{qva_opt}, and \ref{pvtau_opt}. These lines would nearly always intersect the model curves in at least two places for Figs.~\ref{pva_opt} and \ref{qva_opt}, but this would place limits on the possible values of the inclination angle, especially in cases where the optical depth can be estimated from other measurements. Also, if the observed Stokes $q$ parameter were negative, we could say based on Fig.~\ref{qva_opt} that the bow shock was  optically thick and viewed at an inclination angle greater than $90^{\circ}$. With an observed value of $p$, using Fig.~\ref{pvtau_opt} we could constrain the inclination angle if we had spectral information that probed the CSM optical depth, or constrain the optical depth if we had radial and transverse velocity information that limited possible inclination angles. 

\subsection{Thomson Scattering with Absorption}
\label{vara}

In this section, we investigate cases in which the albedo $a$ of the CSM is not unity (that is, at each interaction, photons have a chance of being absorbed rather than scattering). The \textit{SLIP} code can assign a user-specified albedo to the scattering material, but it also has the capability to calculate a self-consistent albedo using the input temperature and optical depth. In our simulations, the CSM is composed of pure hydrogen, both ionized and neutral. Thus, in the case of variable albedo, we assume photons may be absorbed by hydrogen atoms via both bound-free and free-free processes. The resulting absorption opacity is a function of photon wavelength. Although \textit{SLIP} can consider any range of wavelengths, for simplicity we assume a single optical wavelength of 6040 \AA; this represents an intermediate value in the hydrogen opacity curve and avoids absorption edges. With this wavelength, the combinations of temperature and optical depth we consider give rise to albedo values that span the possible range from 0 to 1 (Table~\ref{tab:albedo}). 

When we allow the albedo to vary, we first calculate the hydrogen absorption opacity $\kappa_H$ for 6040 \AA~via Eq.~2 in \citet{wood_96}. Using the ionization fraction $x$ found as above in \S~\ref{methods}, we then set the albedo to be the ratio of scattering to total opacity: $a=0.4x/(0.4x+\kappa_H)$. Because we assume $x$ to be constant throughout the CSM for computational simplicity, $a$ is constant also. Table \ref{tab:albedo} presents the calculated albedo values for different temperatures and optical depths for our assumed wavelength of 6040 \AA. For a given optical depth, the albedo increases with CSM temperature. In the subsections below, we discuss our model predictions of the polarization behaviour as a function of optical depth and temperature when the albedo is allowed to vary. As in the pure-scattering case, position angles for these models are generally near $\Psi=0\degr$.

\begin{table}
\caption{Albedo values calculated by \textit{SLIP} when $a$ is not constrained to be 1, for an assumed wavelength of 6040 \AA~ and different CSM temperatures and reference optical depths (\S~\ref{vara}).}
\label{tab:albedo}
\centering
\begin{tabular}
{|c|c|c|c|c|}
 \hline
$\tau_0$ & 5,000 K & 8,000 K & 10,000 K & 20,000 K\\
 \hline
0.5 & 0.468 & 0.862 & 0.927 & 0.985\\
2.0 & 0.180 & 0.609 & 0.761 & 0.942\\
 \hline
\end{tabular}
\end{table}

\subsubsection{Temperature dependence -- resolved bow shock} \label{result_temp_r}

As the CSM temperature increases, the albedo increases for a constant input optical depth, as shown in Table \ref{tab:albedo}. This causes our results to deviate from the pure Thomson-scattering results (\S~\ref{thomson}), especially at lower temperatures.

\begin{figure*}
\includegraphics[width=\textwidth]{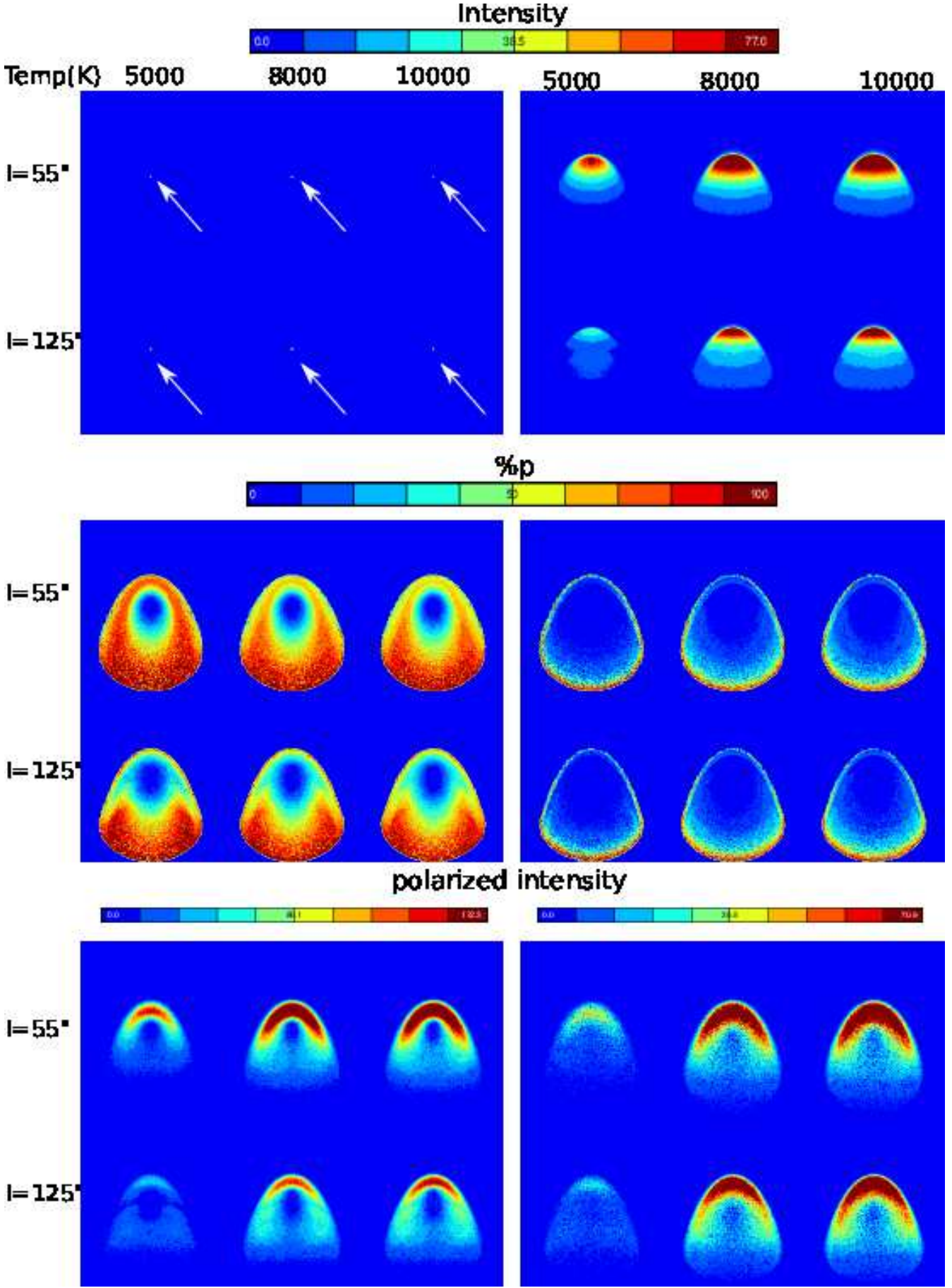} 
\caption{Intensity, polarization, and polarized intensity maps for resolved bow shocks illuminated by a central source (\textit{left}) and the distributed source (\textit{right}) for the case of CSM albedo $a<1$ (\S~\ref{result_temp_r}) and an optical depth of $\tau_0=0.5$. We show two inclination angles symmetric about $90^{\circ}$. CSM temperature increases from left to right in each row. Intensities are in arbitrary units. In the central-source intensity maps, arrows indicate the location of the star. Higher resolution figure can be found \href{http://portfolio.du.edu/hoffman-group/page/66722}{here}}
\label{map_05_difftemp}
\end{figure*}

Fig.~\ref{map_05_difftemp} shows maps of intensity, percent polarization and polarized intensity for two different viewing angles and three different temperatures for $\tau_0 = 0.5$. In the central-source case (\textit{left side}), the scattered intensity is too faint to be seen on this linear scale, as discussed above in \S~\ref{result_opt_r}. In this case we also see little change in polarization as the temperature increases (corresponding to increasing albedo; Table~\ref{tab:albedo}). This is because the overall number of photon interactions is small at this low optical depth.
As in the pure scattering case, the polarization near the bow head is lower for the higher viewing angle. In this case, photons are removed from the beam by absorption in addition to scattering, but the result is the same.
Polarized intensity is concentrated near the bow head as in the pure scattering case; it increases with increasing temperature as the photons undergo more scattering events relative to absorption events, which increases their likelihood of escaping.

In the distributed case (\textit{right side}), there is little variation in polarization with respect to either temperature or viewing angle, again due to the low number of interactions. The polarized intensity maps show a very similar behaviour to those of the central-source case, with more polarized intensity at higher temperatures. 

When absorption is present, the relation between the polarization and polarized intensity maps for central-source and distributed cases is quite similar to that discussed above for the pure-scattering scenario (\S~\ref{result_opt_r}). As we noted there, the difference in polarization maps suggests a possible observational diagnostic for the CSM:star brightness ratio.
By contrast, if we compare the maps including absorption  to the corresponding pure-scattering maps in Fig.~\ref{map_diffopt} (\textit{middle column}), we see very little difference, suggesting that polarization observations may not be able to constrain the albedo of the scattering material in cases of low optical depth. 

\begin{figure*}
\includegraphics[width=\textwidth]{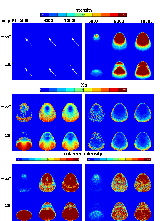} 
\caption{As in Fig.~\ref{map_05_difftemp}, but for $\tau_0 = 2.0$. ``Ringing'' patterns are not physical, but rather due to the discrete model grid (Fig.~\ref{taudens}). Higher resolution figure can be found \href{http://portfolio.du.edu/hoffman-group/page/66722}{here}}
\label{map_2_difftemp}
\end{figure*}

In Fig.~\ref{map_2_difftemp}, we present the intensity, polarization, and polarized intensity maps for the case of variable albedo and an optical depth of $\tau_0 = 2.0$. These maps were created using models with the same number of input photons as Figs.~\ref{map_diffopt} and \ref{map_05_difftemp}, but look grainy because so many of the emitted photons become absorbed in the case of higher optical depth. Because of the relationship between albedo and temperature (Table \ref{tab:albedo}), absorption effects are strongest for $T=5000$ K (\textit{left column of each set}). 

In the central-source case (\textit{left side}), we once again find a very small intensity contribution from scattered light (\S~\ref{result_opt_r}). At lower temperatures, the polarization maps show a ``dark belt'' at mid-latitudes that is not present at higher temperatures. This belt delineates the region of highest optical depth in the CSM, with $\theta$ values slightly less than the cutoff angle (Fig.~\ref{taudens}; Fig.~\ref{bowsketch_los}). In this region, photons that would normally reach the observer via multiple scattering are instead being absorbed. As the temperature increases, photons are again more likely to scatter at each interaction, so the dark belt disappears. At the higher viewing angle, the polarization is highest in the lower portion of the image. This can be attributed to the increased importance of photons backscattering from the CSM interior (Fig.~\ref{bowsketch_los}, cases $c$ and $d$), combined with a lower density in the CSM facing the observer. Like the polarization, the polarized intensity is concentrated towards the lower portion of the image for the higher inclination angle, whereas for the lower angle the polarized intensity is highest near the bow head. These differences are explained by the longer line of sight for higher angles (described in \S~\ref{result_opt_r}), which greatly increases the probability of absorption. Polarized intensity increases with temperature, as expected due to the decreasing importance of absorption at higher temperatures. 

In the distributed case (\textit{right side}), the intensity images for the first time show a significant contribution from the interior of the bow shock at higher inclination angles, as emission from the front side is suppressed by absorption. The polarization is more widely distributed across the shape for lower temperatures, but becomes more concentrated near the edges (similar to the cases of pure scattering and absorption at low optical depth) as temperature increases. At lower temperatures, most of the scattered photons become absorbed and very few escape, making cancellation effects less efficient and allowing a polarization signal to arise from regions other than the edges. At higher temperatures, more scatters increase cancellation and we approach previously considered cases. The polarized intensity maps behave similarly for the distributed case as for the central-source case.

Taken together, Figs.~\ref{map_diffopt}, \ref{map_05_difftemp}, and \ref{map_2_difftemp} suggest that observational constraints on the temperature of the bow shock (in cases where electron scattering dominates) may be possible, but only in cases of higher density/optical depth. For less dense shock structures, the resolved polarization and polarized intensity maps appear similar whether or not absorption is included. However, at higher densities, new features appear when absorption is important, such as the dark belt in polarization and the interior of the shock cone in intensity and polarized intensity. These features could serve as temperature and density indicators in actual observations.

\subsubsection{Temperature dependence -- unresolved bow shock} \label{result_temp_ur}

\begin{figure*}
\includegraphics[width=\textwidth]{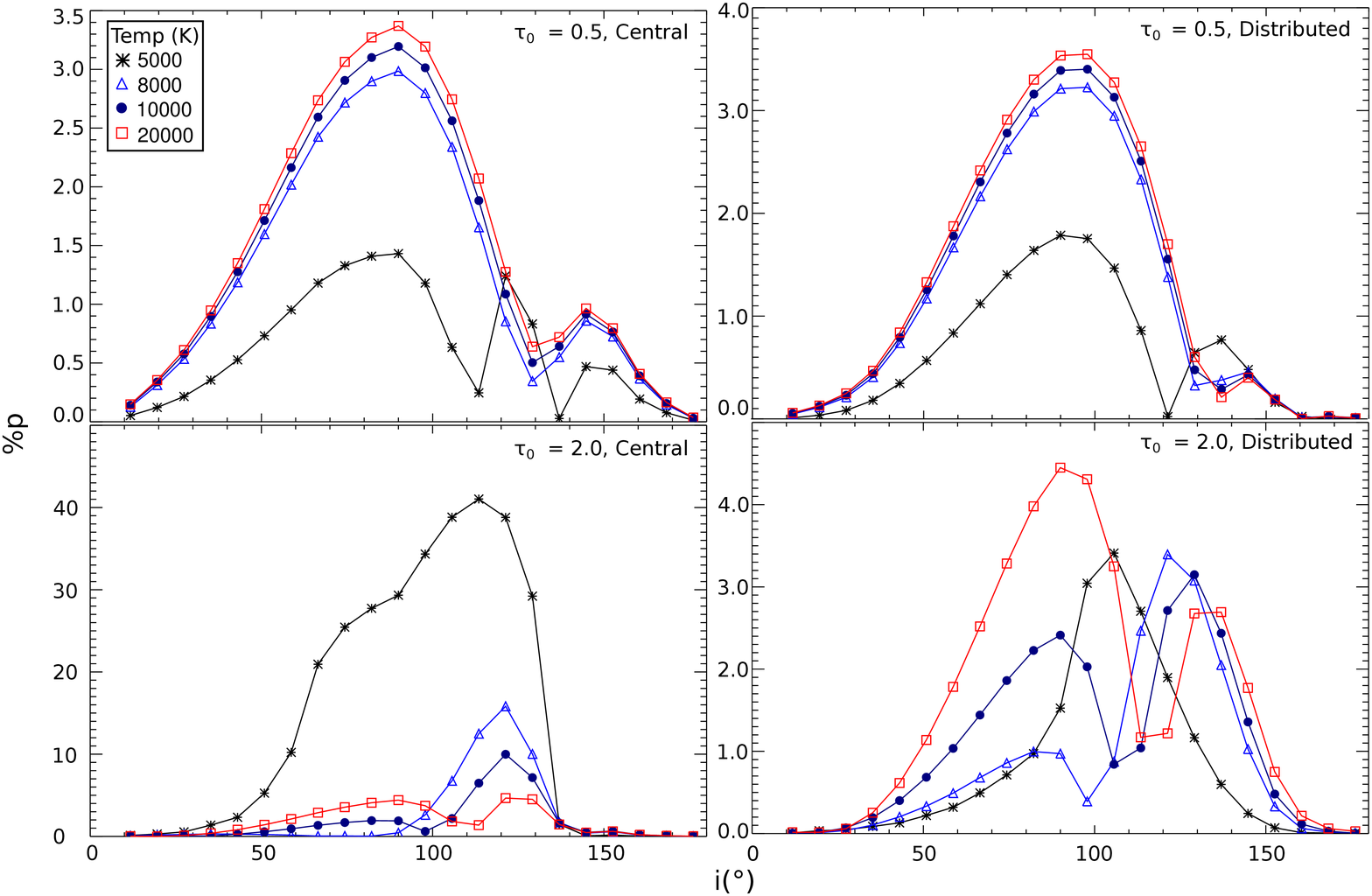}
\caption{Polarization as a function of inclination angle for an unresolved bow shock with different CSM temperatures, for the case of CSM albedo $a<1$ (\S~\ref{result_temp_r}). Photons arise from the central source (\textit{left}) or from the CSM (distributed-source; \textit{right}). Low optical depths are shown in the top row and higher optical depths in the bottom row. Error bars representing $1\sigma$ uncertainties in each model bin (\S~\ref{methods}) are smaller than the plotted symbols.}
\label{pva_difftemp}
\end{figure*}


In Fig.~\ref{pva_difftemp}, we display the polarization variation as a function of viewing angle for models with absorption in the unresolved case, varying both optical depth (\textit{rows}) and temperature (\textit{columns}).
In the lower optical depth regime (\textit{top row}), the increase in albedo with temperature (Table 
\ref{tab:albedo}) causes the degree of polarization to increase at most viewing angles for both central and distributed photon sources. When the albedo is low, photons tend to be absorbed rather than scattered, which lowers the overall degree of polarization (as seen in \citealt{wood_96}). As the albedo increases, photons that have been scattered and thus polarized are more likely to escape the bow shock. Hence we see an increase in polarization for higher temperatures. 

At high optical depths, the albedo is generally small, and increases with increasing temperatures (Table \ref{tab:albedo}). Thus at lower temperatures, only small numbers of photons can escape from the bow shock, and those that escape tend to be highly polarized. As the temperature increases, more photons can escape without scattering; this decreases the overall fractional polarization value. We see these effects in the case of the optically thick CSM illuminated by a central source (Fig.~\ref{pva_difftemp}, \textit{lower left panel}), where polarization values are very high (up to 45\%) and the peak near $90\degr$ is suppressed for all temperatures. There is a prominent second peak near $i=130\degr$; as the temperature increases, the degree of polarization decreases at this higher viewing angle. 
We attribute the suppression of the $90\degr$ peak to the combination of higher optical depths and lower albedos, which together increase the chance for a photon to be absorbed. Inspection of the flux characteristics of these models shows that  most of the photons escape in the wings of the bow shock, where the optical depth is lower due to our cutoff angle. Thus the secondary peak we discussed in the pure-scattering case (\S~\ref{result_opt_ur}) dominates the polarization in these models.

The secondary peak is also prominent in the optically thick, distributed-source cases (\textit{lower right}), although the polarization values are smaller than for the central-source models because more photons escape directly from near the surface of the CSM. 
The $90\degr$ peak is still present for most temperatures. At $T=5000$ K, however, only the secondary peak contributes, while the $90\degr$ peak is completely suppressed by absorption (Fig.~\ref{map_2_difftemp}, right-hand side). The polarization is almost entirely due to photons arising and scattering near  the interior surface of the CSM. Because very little polarized intensity arises from the outer surface, in this extreme scenario the secondary peak shifts to a viewing angle of $\approx 110\degr$, at which the interior first begins to be visible.
In the high-density cases for both photon sources, the models with the highest temperatures approach the behaviour of the pure scattering case as $a\rightarrow 1$. 

\subsubsection{Optical depth dependence -- resolved bow shock} \label{result_opt_r_vara}

Using Figs.~\ref{map_05_difftemp} and \ref{map_2_difftemp}, we can also assess our resolved results as a function of optical depth. The intensity maps vary significantly with optical depth in the case of the distributed source. At the higher inclination angle, the intensity is concentrated near the bow head for $\tau_0 = 0.5$, whereas for $\tau_0 = 2.0$ the intensity arises primarily from the wings and interior of the bow shock structure. 

For all temperatures, the degree of polarization decreases with increasing optical depth. We attribute this behaviour to the decrease in albedo with $\tau_0$ shown in Table \ref{tab:albedo}. 
For the central source at the lower temperature of $5000$ K, the ``dark belt'' effect occurs for higher optical depths only, due to a lower albedo combined with increased photon interactions.
For the distributed source, the polarization is primarily concentrated near the edges as in the pure scattering case. However, in the lower-temperature case viewed from $i = 125\degr$, some polarization arises from the upper portion of the bow shock for $\tau_0=2.0$, which is not seen at $\tau_0=0.5$. This occurs because when absorption is frequent, cancellation of Stokes vectors cannot happen for $\tau_0=2.0$ as efficiently as in the case of $\tau_0=0.5$, so some net polarization remains.

In polarized intensity, the two optical depths produce very different maps. For the central-source case, at $\tau_0 = 0.5$ the polarized intensity is concentrated near the bow head for both viewing angles, while for $\tau_0 = 2.0$ at the higher viewing angle, the polarized intensity is concentrated towards the lower portion. This is because when the density near the bow head is high and $a<1$, photons have a better chance of being absorbed in those regions. In the lower portion of the map, for $\theta$ values greater than the cutoff angle, the density is much lower; thus most of the photons that are polarized can escape the bow shock. These photons arise primarily from the interior of the shock cone, which is visible at the higher angle. 
We see a similar effect in the distributed-source case.

Because of these optical depth variations, observed polarized intensity maps can potentially constrain the optical depth of the bow shock material as well as the structure's inclination angle. Comparison of observed maps with these predictions can also help identify the source of illumination and thus relative brightnesses of star and CSM, as discussed above (\S~\ref{result_opt_r}).

\subsubsection{Optical depth dependence -- unresolved bow shock} \label{result_opt_ur}

We can isolate optical depth-dependent behaviour for unresolved cases by comparing top to bottom panels in Fig.~\ref{pva_difftemp}. For a constant temperature, the location of the polarization peak is different for the two optical depths. In the optically thin case, the peak is near $90\degr$ \citep[as predicted by analytic models, e.g.][]{brown_1977} for both the central and distributed cases. In the optically thick case, the peak shifts to higher inclination angles for both photon sources. For a constant temperature, increasing optical depth leads to decreasing albedo. Thus, when $\tau_0$ is high, very few photons can escape from  the denser central regions of the bow shock. Instead they escape from higher viewing angles, giving rise to the secondary peaks for higher optical depth. 

In the central-source case, the model with $T=5000$ K and high optical depth produces the highest polarization in any of our models, because it has the lowest albedo. As discussed in Section \ref{result_temp_ur}, this scenario results in a low number of escaping photons (mainly those scattering from the interior surface) and thus high polarization magnitudes. At $90\degr$, instead of a polarization peak, this extreme case shows a small ``notch'' that we attribute to the prominence of the ``dark belt'' discussed in  \S~\ref{result_temp_r}: at edge-on inclinations, this belt will dominate the polarization signal, with very few photons escaping from either the bow head or the interior.


In the distributed-source case, the models evolve from single-peaked to a double-peaked shapes as $\tau_0$ increases. At higher optical depths, the $90\degr$ peak is suppressed and the secondary peak begins to dominate, due to the fact that scattered photons can more easily escape at higher inclinations once absorption is present. At the lowest temperature, for which the albedo is close to 0, the $90\degr$ peak completely disappears and the polarization is due entirely to photons arising and scattering near the interior surface of the CSM (\S~\ref{result_temp_ur}).


\section{Observational implications}
\label{obsimp}

We close by discussing potential observational implications of the electron-scattering results presented here (subject to the model limitations discussed below in \S~\ref{conclusion}). These are useful as limiting cases and to lay the groundwork for future models that will include both electrons and dust as polarizing mechanisms.

In the case of a resolved bow shock, detailed polarization maps are rare in the literature, so it is not currently possible to compare our image predictions with actual observations. (The observations by \citealt{2013A&A...551A..35R} provide a notable exception, but these authors observed a known dusty source and obtained only 9 polarization measurements across the bow shock.) Our results show that in future observational efforts, both polarization and polarized intensity maps may provide useful diagnostics. Polarization maps are relatively insensitive to viewing angle except in the case where absorption is significant (Fig.~\ref{map_2_difftemp}). However, because the differences between central- and distributed-source models are greatest in polarization (Figs.~\ref{map_diffopt}, \ref{map_05_difftemp}, and \ref{map_2_difftemp}), these maps may provide information about the relative brightnesses of source and bow shock. This could lead to more realistic models for individual stars that consider both central and distributed photon sources (\S~\ref{conclusion}). Polarization maps can also reveal information about the temperature of the bow shock when absorption is important. In particular, an observed ``dark belt'' (Fig.~\ref{map_2_difftemp}) would indicate a relatively low CSM temperature and high density. Polarized intensity maps can distinguish between two symmetric viewing angles in the case of higher optical depths (Figs.~\ref{map_diffopt} and \ref{map_2_difftemp}). Although we have not presented them here, \textit{SLIP} can also produce position angle maps for comparison with observations. The position angles in our models are consistently $\approx0\degr$ for most viewing angles, but flip to near $90\degr$ at high inclinations and optical depths when $q$ is negative.

For unresolved bow shocks (or cases in which a bow shock is predicted to exist, e.g. \citealt{Neilson_Ignace_Smith_Henson_Adams_2014}), we measure a single polarization value corresponding to a single viewing angle. This corresponds to a horizontal line in figures such as Figs.~\ref{pva_opt}, \ref{qva_opt}, \ref{pvtau_opt}, and \ref{pva_difftemp}. If interstellar polarization can reliably be removed, this could place constraints on the viewing angle if optical depth can be estimated (Figs.~~\ref{pva_opt} and \ref{qva_opt}), or vice versa (Fig.~\ref{pvtau_opt}). A measurement of a negative value of Stokes $q$ (accounting for the orientation of the bow shock on the sky, e.g. using the proper motion of the star) would provide a particularly strong viewing angle constraint (Fig.~\ref{qva_opt}). Finally, a polarization measurement compared with the curves in Fig.~\ref{pva_difftemp} could provide constraints on the CSM temperature, particularly at low optical depths or for centrally-illuminated shocks.

\section{Conclusions and future work}
\label{conclusion}

We investigated the polarization arising from electron scattering within an idealized stellar wind bow shock, for cases of illumination by a central star and self-illumination by the shock region. We studied how different parameters impacted the polarization behaviour for both pure scattering and scattering with absorption cases. As expected, polarization is highly dependent on viewing angle for all models. Multiple scattering significantly modifies the behaviour of the polarization with respect to analytical predictions assuming single scattering. For very low optical depths, our simulations reproduce the analytical $\sin^2 i$ dependence of \citet{brown_1977}, but many of our models show a secondary peak at higher inclination angles attributable to increased $-q$ polarization caused by multiple scattering.

In the case of pure scattering (albedo $a=1$), we find that the optical depth of the bow shock significantly affects the resulting polarization behaviour, while its temperature does not. In addition, while changing the photon source (light arising from the central star vs. from within the bow shock) does not drastically modify the polarization curves for the unresolved case, it does change the appearance of the polarization and polarized intensity maps for resolved bow shocks. We have presented the central- and distributed-source cases separately here for clarity, but typically both should contribute simultaneously to the observed polarization. \textit{SLIP} has the capability to combine the two cases by specifying the relative brightnesses of the star and CSM; we will investigate these cases in the future when modeling particular bow shocks.

When the albedo is not fixed at 1, but instead calculated using input parameters, we find that the polarization depends both on temperature and optical depth. In this case, absorption effects cause dramatic departures from $\sin^2 i$ behaviour, particularly for higher optical depths and lower temperatures. These effects also produce resolved polarization maps that differ from those of the pure-scattering and low optical depth cases. We have chosen a representative optical wavelength of 6040 \AA~to represent these cases, but this can be changed to correspond to specific observed scenarios.

We made several simplifying assumptions in creating these models, which should be kept in mind when interpreting the results. First, we chose a specific value of $\alpha={V_*}/{V_w}=0.1$ to correspond to winds from hot stars (\S~\ref{results}). For cooler stars, $\alpha$ will be larger, and this will increase the density of the bow shock via Eq.~\ref{rho} \citep[see also Fig.~4 of ][]{Wilkin_1996}. Thus, we expect that the results for cooler stars will be similar to those of the high optical-depth cases we discuss here. 

We also chose a specific standoff radius $R_0$ (\S~\ref{results}) for consistency in the models presented here. In the pure-scattering case, polarization behaviour does not depend on $R_0$, but for the more realistic case of variable albedo, the polarization may differ from the results presented here. This is due to the way we defined the thickness and density of the \citet{Wilkin_1996} bow shock, as discussed in \S~\ref{results}. A study investigating the use of polarization as a diagnostic of the stellar mass-loss rate or ISM density would need to assume or measure a value for $R_0$ in order to generate models with the appropriate CSM opacity and albedo. Such a study could be undertaken with \textit{SLIP}, but is beyond the scope of this paper because of the wide range of possible $R_0$ values. In Paper II, we plan to compare \textit{SLIP} models with polarization measurements of bow-shock sources with measured $R_0$ values, and will adjust the models accordingly.

We have not investigated the effect of ionized stellar wind material filling the interior of the bow shock, but we expect this would decrease the overall polarization magnitude without significantly affecting its behaviour as a function of viewing angle (particularly in the case of photons arising from the central source). We will explore the polarization contributions of interior scattering material in Paper II.


We also note that the bow shock solution presented by \citet{Wilkin_1996} is an idealization that assumes a stable and highly evolved bow shock, as shown by hydrodynamic models \citep{Mohamed_2012}. Resolved polarization or polarized intensity maps that show bow shock shapes similar to those in our models would thus provide information about the age of the observed bow shock, which in turn can reveal the evolutionary state of the star, as discussed in \citet{Mohamed_2012}. Younger bow shocks or bow shocks with instabilities due to a high-density region of the ISM \citep{Meyer_2014} or a star moving with a high space velocity \citep{Meyer_2015} will show different morphologies than the idealized shape considered here. We expect these cases will display broadly similar polarization features, but detailed studies will require additional modeling. We plan to investigate clumpy shock structures in a future contribution.

We recognize that dust scattering is an important contributor to the observed polarization of actual bow shocks that we have not treated here. In fact, most observations of stellar wind bow shocks have been obtained using IR data \citep[e.g.,][]{Kobulnicky_2016,Ueta_2006,Ueta_Izumiura_Yamamura_Nakada_Matsuura_Ita_Tanab_Fukushi_2014,Peri_Benaglia_Brookes_Stevens_Isequilla_2011}. The \textit{SLIP} code can treat dust scattering, and we will investigate its behaviour in Paper II. 
We will discuss the variation in polarization behaviour at different wavelengths as well as for different dust grain models.

\section{Acknowledgements} 
We thank Dr. D. Meyer, Dr. T. Ueta, and our anonymous referee for thoughtful comments that have greatly improved the paper. We also thank Dr. B. Whitney for helpful code consultations. This work has been supported by NSF awards AST-0807477 and AST-1210372 and by a Sigma Xi Grant-In-Aid of Research.

\newpage
\bibliographystyle{mnras}
\bibliography{electron_new.bib}

\appendix
\section{Calculation of \lowercase{$b(\theta)$}}
\label{appendix}
\begin{figure}
\includegraphics[width=\columnwidth]{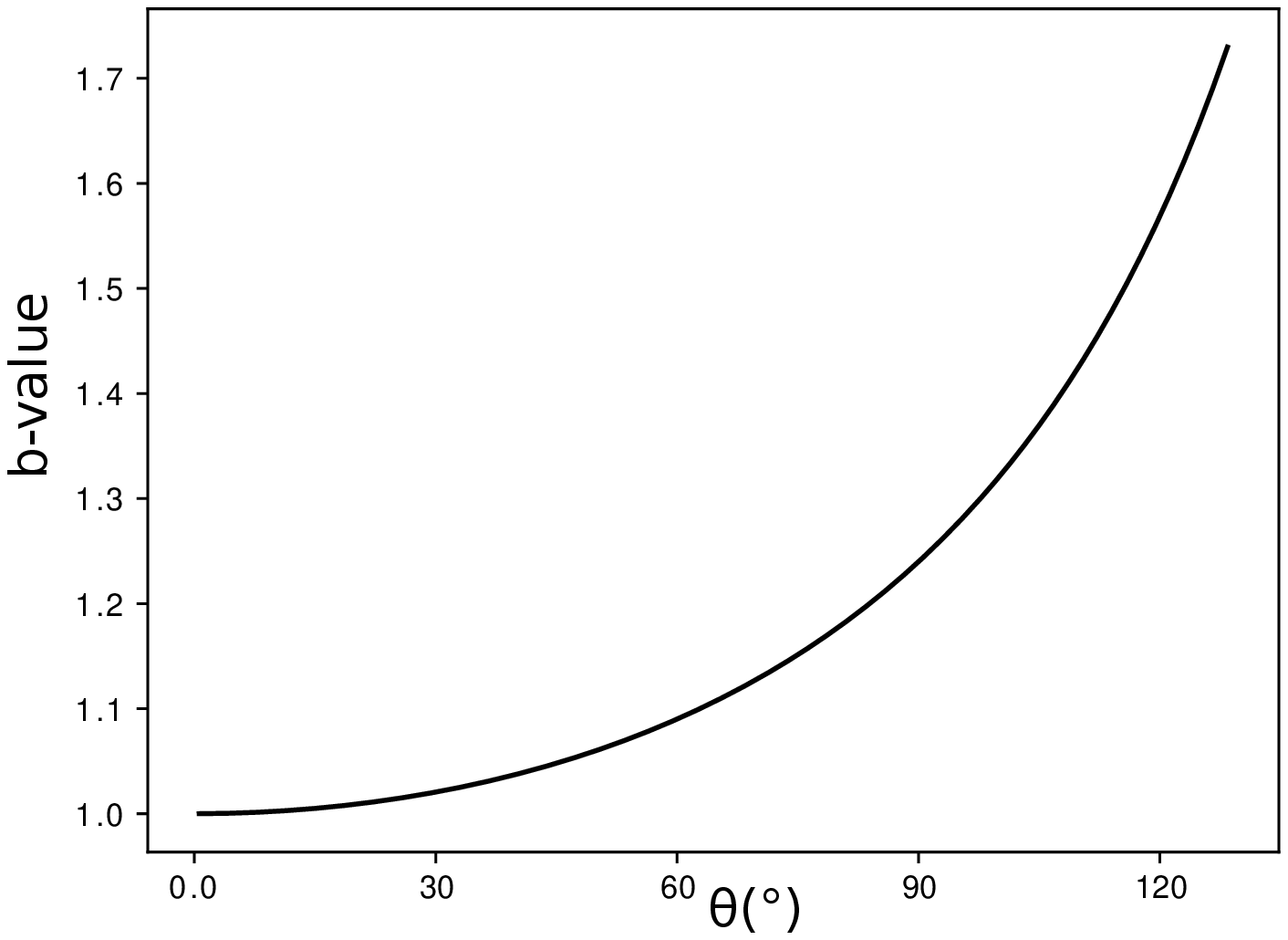}
\caption{Variation of the $b$ factor with $\theta$.}
 \label{gvtau}
\end{figure}

The $b$ factor referred in section \ref{methods} is given by
\begin{equation}
b(\theta) = \sqrt[]{1+\frac{1}{4} \bigg(\frac{\theta \csc \theta^2-3\cot\theta+2 \theta \cot\theta^2 }{1-\theta \cot\theta}\bigg)^2};
\label{g-factor}
\end{equation}

\noindent
we show its functional form in Fig.~\ref{gvtau}. This factor arises from the arc length formula involved in the calculation of surface area. Its presence here is due to the fact that an area element of the bow shock is not generally oriented normal to a radial vector from the star, with respect to which we define the optical depth $\tau$.
The bow shock is axisymmetric and therefore can be considered a surface of revolution about the $z$-axis. The surface area, $S$, is defined in terms of the curve described by the bow shock at a fixed azimuth. The path length of the curve from the bow head to some point downstream along the shock at position $(r,\theta)$ is represented by $l$. The surface area for that portion of the bow shock is then

\begin{equation}
S= \int 2 \pi\, r\, \sin \theta \, dl,
\end{equation}

\noindent where $r$ is the radius from the star to the curve, and $dl$ is given by

\begin{equation}
dl = \sqrt[]{r^2+\bigg(\frac{dr}{d\theta}\bigg)^2} \,d\theta.
\end{equation}
 
After substituting the expression for $dl$ into the $S$ integral and factoring, we find the surface area becomes

\begin{equation}
S= \int 2 \pi\,r^2 \, \sin \theta\, d\theta \, \sqrt[]{1+\bigg(\frac{d\ln r}{d\theta} \bigg)^2}.
\end{equation} 

The term under the square root is what we call the $b$ factor. Thus,

\begin{equation}
b(\theta) = \sqrt[]{1+\bigg(\frac{d\ln r}{d\theta} \bigg)^2},
\end{equation}
\label{gderiv}

\noindent where $r$ is given by Equation~\ref{radwill} for the bow shock. Putting Equation~\ref{radwill} into Equation~\ref{gderiv} and simplifying, we obtain Equation~\ref{g-factor}. In the code, we implement this factor discretely by calculating $b$ for each grid cell.

\bsp	
\label{lastpage}
\end{document}